\pdfoutput=1

		\documentclass{isprs}
		\usepackage{hyperref}
		\usepackage{subfigure} 

		\usepackage[utf8]{inputenc}
	
		\usepackage{varioref} 
	 
		\usepackage{natbib} 
	\usepackage{enumitem}
		\usepackage{graphicx}
		\DeclareGraphicsExtensions{.pdf,.jpg,.png}

		\usepackage[usenames,dvipsnames]{color}
	\usepackage{amsmath}
		\usepackage{algorithm2e}
		\usepackage{microtype}  
		\usepackage{SIunits} 
		\usepackage{listings}
		\usepackage{tabularx} 
		\usepackage{setspace}
		 \DeclareMathOperator*{\argmin}{arg\,min}
		
		\newcommand{\myimage}[3]
					{
					\begin{figure} [h!]
						\begin{center}
							\includegraphics[width=\linewidth,keepaspectratio]{#1}
							\caption{#2}  
							\label{#3}
							\end{center}
					\end{figure} 
					}

		\newcommand{\myimageHL}[4]
		{
			\begin{figure} [ht!]
				\begin{center}
					\includegraphics[width= #4 \linewidth ,keepaspectratio]{#1}
					\caption{#2}  
					\label{#3}
				\end{center}
			\end{figure} 
		}

		\newcommand{\myimageFullPageWidth}[3]
								{
								\begin{figure*}[ht]
									\begin{center}
										\includegraphics[width=\textwidth,keepaspectratio ]{#1}
										\caption{#2}  
										\label{#3} 
										\end{center}
								\end{figure*} 
								}

	\usepackage{rotating}

\title{User assisted and automatic inverse procedural road modelling at the city scale}
\author{Rémi Cura  $^{A}$, Julien Perret $^A$, Nicolas Paparoditis  $^A$}

\address{ $^A$  Université Paris-Est, IGN, SRIG, COGIT \& MATIS, 73 avenue de Paris, 94160 Saint Mandé, France\\
	first\_name.last\_name@ign.fr
	}

\begin{document}
 



\abstract{	
Cities are structured by roads. Having up to date and detailed maps of these is thus an important challenge for urban planing,
civil engineering and transportation.
Those maps are traditionally created manually, which represents a massive amount of work, and may discard recent or temporary changes.
For these reasons, automated map building has been a long time goal,
either for road network reconstruction or for local road surface reconstruction from low level observations.
In this work, we navigate between these two goals.
Starting from an approximate road axis (+ width) network as a simple road modelling, 
we propose to use observations of street features and optimisation to improve the coarse model.
Observations are generic, and as such, can be derived from various data, such as aerial images, street images and street Lidar, other GIS data, and complementary user input.

Starting from an initial road modelling which is at a median distance of $1.5 \metre$ from the sidewalk ground-truth,
our method has the potential to robustly optimise the road modelling so the median distance reaches $0.45 \metre$ fully automatically,
with better results possible using user inputs and/or more precise observations.
The robust non linear least square optimisation used is extremely fast, with computing time from few minutes (whole Paris) to less than a second for a few blocks.

The proposed approach is simple, very fast and produces a credible road model.
These promising results open the way to various applications, such as integration in an interactive framework, or integration in a more powerful optimisation method, which would be able to further segment road network and use more complex road model.
}

\maketitle 


\myimageFullPageWidth{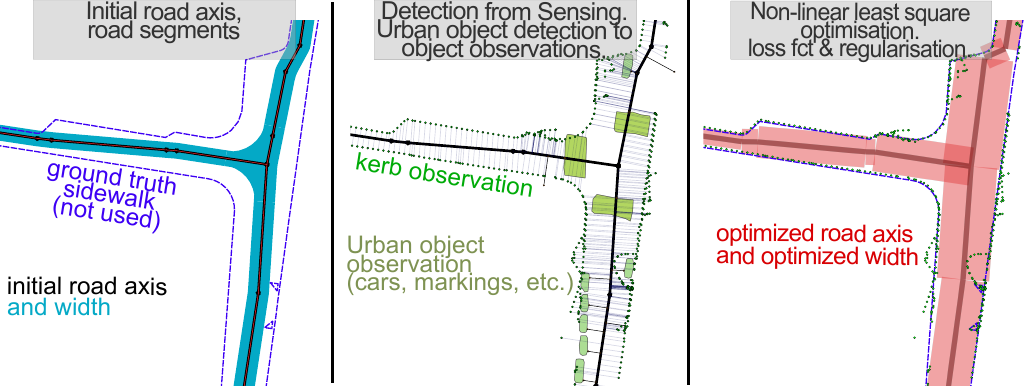}{Approximate road axis network and road width are available, forming a basic road modelling. Various sensing methods produce urban feature detections which are processed into consolidated observations, and assigned to road axis segments. A robust non linear least square optimisation then fits the road modelling to observations. The result is much closer to ground truth, even when the road model is too simple for the actual road configuration (varying width, asymmetric changes, curves). The user can further input observations if necessary.}{optim.fig.graphical_abstract}

%

\section{Introduction}   
\subsection{Problem}

Paris is a large city, with thousands of kilometres of streets.
Unlike highways, the vast majority of roads in those streets do not follow strict design guidelines due to historical reasons (they also pre-date civil engineering guidelines, which have also evolved anyway).
Yet, an up-to-date precise map of those roads is essential for many applications, like city planning, urbanism, traffic analysis, autonomous driving or simply help wheelchairs or stroller users navigate urban space.
Streets are also changing frequently, 
with very frequent public work and planning efforts, 
as well as effects from other civil works and maintenance.
As a testimony, the ground truth data we use dates from 2011, however, actual sensing from 2014 showed that a part of the side-walks had been modified,
and we only looked at a small area of a few blocks (See Fig. \vref{optim.fig.area_of_experiments}, right)!

Manually creating and updating those road modelling is extremely time consuming, 
and we wonder how much time and efforts went into the 1860 Paris plan that include side-walks (Fig. \ref{optim.fig:plan_parcellaire_paris}). 
\myimageHL{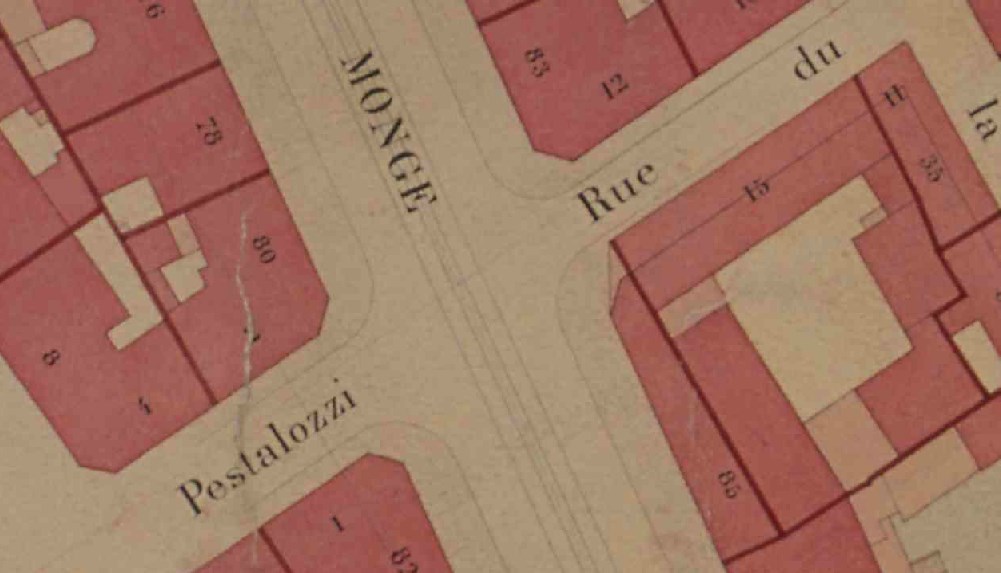}{Plan parcellaire municipal de Paris, 1860. Copyright Archive de Paris.}{optim.fig:plan_parcellaire_paris}{1}
Thus, a mostly automatic method is needed to create/update these road modelling.
Automatic methods relies heavily on data: actual observations that can be leveraged to find an adequate road modelling.
The ways to observe roads are numerous,
be it through street images or street Lidar, aerial images, legacy GIS database or digitized legacy maps.
All those observations sources should be usable by the method, as each could suit a particular situation.
For instance high buildings reduce the interest of aerial images, but have no impact on street Lidar.
Yet street Lidar data is unlikely to be available outside of major cities.

Thankfully, such a road modelling is usually not to be created from scratch.
At least, some kind of road axis network is usually available, either from a database,
or reconstructed from sensing data (aerial image/Lidar, vehicle trajectory, phone tracking, etc.),
although reconstructing a road axis network is already a very challenging task in itself.

Of course observations from sensing are not perfect, especially when several sensing sources are mixed.
Therefore, to model a road is in fact akin to finding the optimal road modelling that fits those observations.
A road model with parameters must be defined, then, a suitable optimisation method can be used to find the optimal values of these parameters.
The observation may be erroneous, and so can be the road axis network.
Finding the optimal road modelling may then not only involve finding the optimal parameter values of the road model,
but also to find the optimal number of parameters,
as well as the observations that should be used.

Moreover, urban roads are sometimes so complex that some form of user interaction is most likely necessary.

\subsection{Related work}  

Road surface reconstruction is a popular topic with widely varying methods depending on the input data, the complexity of expected result, and available computing time. Methods also vary depending if the goal is to reconstruct local road surface or the road surface as part of a road network.

\paragraph{Observation oriented (sensing) for local reconstruction}
Many methods focus on fast methods producing low complexity local road models. This topic is of particular importance for autonomous vehicles, as they need to be able to reconstruct the surrounding roads to be able to navigate properly  . \citep{BarHillel2012} propose a state of the art of (local) road and lane reconstruction for autonomous driving.

Those bottom up methods can use street Lidar as in \cite{Zhang2010b}, street videos  (\cite{Zhang2009a}), or rasterized Lidar (\cite{Yang2013a}).
In those cases, the goal is more to classify which pixels or points pertain to road, rather than reconstruct a high level road model. 

The local reconstruction can also be based on aerial image, such as in road followign methods, which are often used to help an userr, such as in \citep{McKeown1988}, which produces a centerline and approximate road width, and in \citep{Airault1994}.

\paragraph{Observation oriented (sensing) for global reconstruction}
Other methods also work on low level input, but try to reconstruct the road network.
The base input can be for instance GPS traces of vehicles such as in \citep{Roeth2016}, who find then the optimal road network matching these traces with RJ-MCMC. Note that they reconstruct the raod axis network and not the road geometry.
Another input that can be used is radar (SAR), such as in \citep{Tupin1998} where a first local detection step is followed by the global network reconstruction with Markov Random Field.
This workflow is similar to \cite{Montoya-Zegarra2014}, who use learning (local) then Conditional Random Field to reconstruct the road network in urban environment using aerial rasters.

\cite{Fischler1981} method is also based on aerial images. They use multiple road detector and dynamic programming to reconstruct the road surface of the road network. 
\cite{Baumgartner1999} focus on multiscale road extraction and fill gap with active contours. 
They also use context.  
\cite{Ziems2007} use available GIS database information for contextual information and to help learning. 
When the approximate road axis network is available, the problem is then quite different, as demonstrated by \cite{Ravanbakhsh2008} who use local image processing and active contours to precise road delimitation in intersections. 

\paragraph{Model oriented (GIS)}
On the other side of the spectrum, numerous methods focus on rebuilding the main component of a road model: the road axis.
Reconstructing road axis involves topology, and can be performed with various data, such as aerial image (\cite{Montoya-Zegarra2014}), Lidar (see \cite{Quackenbush2013}  state of the art) or vehicle traces (GPS data, see \cite{Ahmed2014}).

Such methods focus on network reconstruction, and usually do not consider road geometry,
with the exception of \cite{Zhang2010a} that reconstruct a road network from GPS traces and also estimate road width,
and \cite{Clode2007}, which use aerial lidar data to reconstruct a road network topology along with the border of the road (width is varying along the road).

\paragraph{Fusion}
Taking advantage of both observation-oriented and model-oriented methods,
some methods fuse low level observation with high level road network.
For instance \cite{Hatger2003} initialise a road segmentation method starting with a road axis network database, and using aerial Lidar. Road width and slope is then extracted from the segmentation, but the road axis are constant.
Similarly, \cite{Boyko2011} start from an approximate urban road network and use it to initialise a road surface seeking method (snake) based on Lidar raster. The road surface is the final result.

Both these methods directly use low level data (Lidar) to segment road surface.

\paragraph{This work}
In this work we take a different approach because we do not use the same input data.
First we do not directly use raw sensing data. Instead, we rely on other methods that extract semantic information about urban features from these raw data (for instance, markings, cars, signs, etc.).
These methods may work on several type of raw sensing data such as aerial images, street Lidar and images, other GIS data, and user input.
 
Second we start from an already existing rough road model composed of a road axis network where each raod axis as an approximate road width.

Our goal is then to adapt this rough road model to the observations we have.
To this end, we associate the semantic observations extracted from low level detections to the road model. 
We then optimise the road model so it better fits these observations.
The result is a fitted high level road model.

\subsection{Approach}

Road modelling as a whole is complex, especially if we want to be able to use diverse observation type and do so fast enough to allow user interaction.
Moreover, even a superficial look at Paris streets outlines numerous odd roads configurations that would be complex to extensively integrate into a road model, if possible at all.
A complex model and diverse uncertain observations make a dangerous mix.
Therefore, we chose to model roads in a very simple yet flexible way: a road axis network (composed of road segments) with a road width for each segment.
Such a model is a simple basis upon which more complex cases can be added when necessary.

Then, we break the optimisation problem of finding the optimal road axis and road width given a set of observations in two parts.
The first part aims at robustly finding optimal values of this road model parameters given observations, road axis segments and associated width.
We consider it to be the core and maybe the task that potentially needs the less user intervention.
As such, it should be fast and robust, yielding good results in most situations.
Some areas may need to be manually corrected when the road model is not sufficient to handle a complex situation.

The second part aims at finding the optimal number of road segments, and which observations affect  which segments.
This involves splitting/merging/creating/deleting road segments, and removing  observations.
In our work this part is performed manually. 
It could be automatized using a powerful optimisation framework, which we leave for future work.
 
\subsection{Plan}  
The rest of this chapter is structured as follows:
In Section~\ref{optim.method}, we justify the choice of our road model and optimisation method.
Then, we explain how we create observations from sensing detection and how we use them in the optimisation process.
In Section~\ref{optim:result}, we show results of observation creation and usage in the optimisation, with optimisation results for various observations and situations. 
In Section~\ref{optim.discussion} we discuss results, limitations and perspectives.



\section{Method}
\label{optim.method}
\subsection{Choosing a model and optimisation method}
	\subsubsection{Context}
	It seems that the way a problem is modelled strongly depends on the optimisation method that will be used.
	Therefore, before exposing our modelling and optimisation method, we explain our goal and the context.
	
	We suppose we have access to a road axis network with associated estimated road width.
	As explained in \cite{Cura2015a}, this hypothesis is not a reducing one, as those data are fairly accessible and when not available can be estimated from various inputs (aerial images, LIDAR, GPS, etc.).
	
	These road axis and road width are only approximate, and lead to a first credible roadway modelling solution, but the goal is to better estimate these values.
	For this, we suppose we also have access to observations related to road or street objects (in a broad interpretation), for instance road markings, kerbs (separation between roadway and side-walk), public lights, etc.
	Again, this hypothesis is not very limiting, as many methods exist to extract those observations from images, aerial view, Lidar, digitized maps, etc. \\
	We interpret an observation as the probable presence of an object, as such each observation is completed by a confidence measure.
	Each object type has a defined behaviour regarding road surface. For instance, pedestrian crossings are expected to be \emph{within} the roadway (road surface).
	From an optimisation perspective, the goal is then to find the best parameters of the model to fit the observations.

	\subsubsection{Optimisation requirement and choice}
	Following the capabilities of StreetGen (\cite{Cura2015a}), we seek a method that can work seamlessly from street to city scale.
	Moreover, because observations may be missing entirely in some place, or be noisy, the method must allow seamless user input integration, as well as user override.
	More than user input integration, the method also has to be fast enough to be interactive at street scale so that user interaction is possible.
	Lastly, observations can concern many different street objects, and take many forms depending on the sensing method.
	Therefore, the optimisation method has to be generic enough to integrate many types of information about urban objects.
	
	We considered optimising on loop of kerbs around a city block, but we rejected it because the street network could be incomplete, and because optimising one isolated street should be possible.
	
	We choose to formalize the problem as a mechanical problem, where observations exercise forces over the road axis, and road axe are also subject to forces that resist changes so the final solution is not too far from the initial one.

\subsection{Modelling the problem}

	\subsubsection{Model}
	We consider the road axis (polylines constituted of road segments) as a set of connected points.
	To each road segment is associated a road width value (parameter / variable).
	Segment and width together implicitly describe a simple roadway (road surface), that is a set of rectangles.
	The variables (i.e. the only values that will be changed by optimisation, which we will call parameters) of the optimisation are then 3D points (3 scalars per points) constituting the road axis vertex, and for each road segment, the road width (on scalar per segment).
	See Figure \ref{optim.fig.optimisation_model} and equation \vref{optim.eq.global_optim}.

	\myimageHL{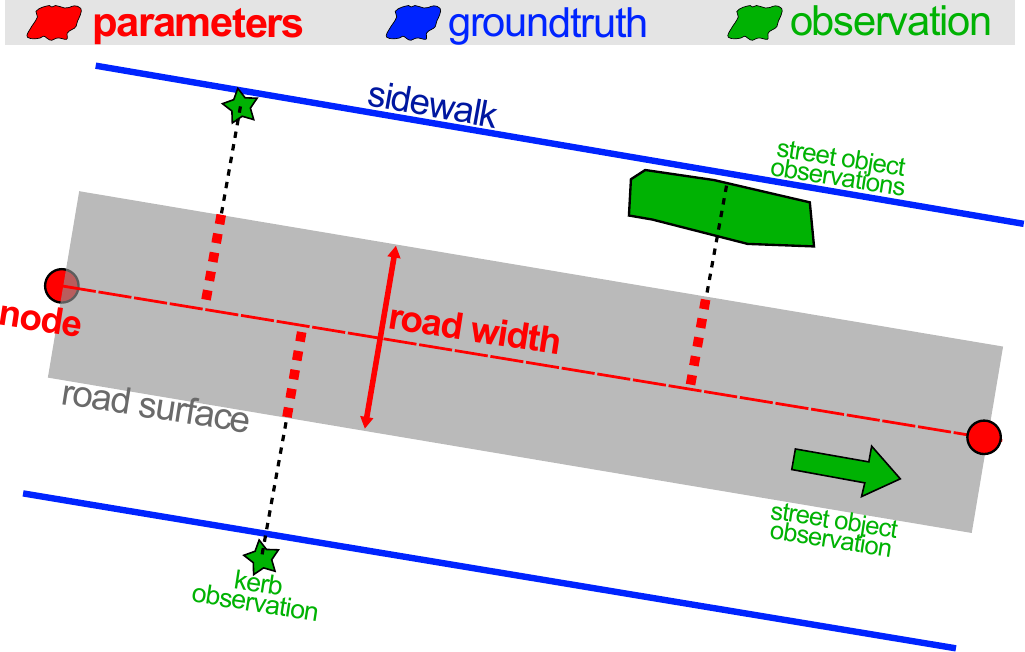}{A roadway is modelled with connected nodes and width, which implicitly describes a crude road surface (a rectangles). Observations are defined regarding the implicit road surface.}{optim.fig.optimisation_model}{1}
	Observations are semantic 2D geometries (points, polylines, polygons, geometry collections) associated with a confidence.
	
	The semantic of street objects is their type (about 100 classes, such as car, street furniture, building, etc.).
	Each street object class is associated with an overall precision, confidence, class weight,
	as well as the expected class behaviour regarding road way.
	We consider those as settings of the optimisation, as those depend on data and knowledge on road.
	Each class has an expected position regarding roadway, which is either \emph{within} the roadway area or \emph{outside}. 
	The expected position can also be defined as in the border of the road area (in or out of the road), or undefined.
	Furthermore, if the class is defined as being on the border of the roadway, it has an additional expected distance to the border (See Fig. \ref{optim.fig.border_in_out}).
	
	\myimage{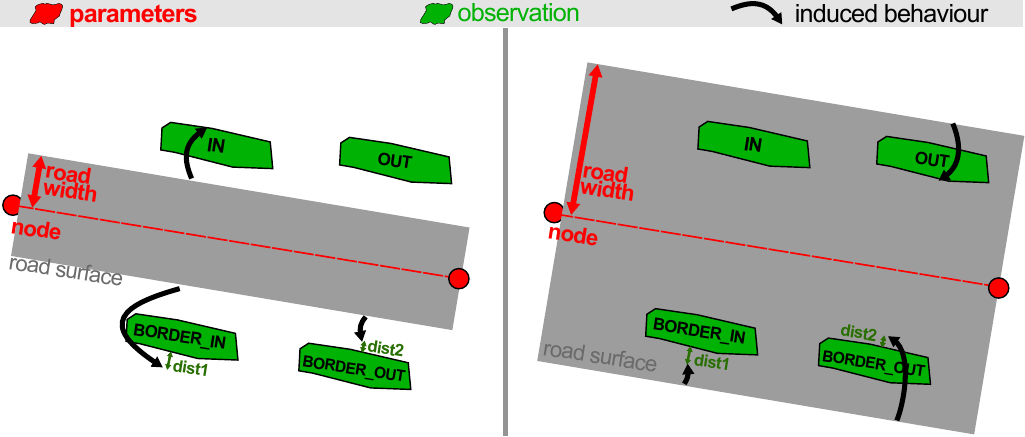}{Each expected observation position is defined regarding the implicit road surface, with four options: In, Out, Border\_in, Border\_out. When Border is used, a distance to the road surface limit can be defined (dist1, dist2 here).}{optim.fig.border_in_out} 
	
	\subsubsection{Optimisation method}

	We choose to use a robust non-linear least square optimisation to find the optimal road axis and road width.
	We use a generic open source tool to do so (\cite{Agarwal2016}).
	The main reasons to choose this family of optimisation is that given observations and parameters (a road axis and road width),
	we can significantly measure how well the parameters fit the observations, and more importantly, we can explicitly compute changes on parameters to improve fitting.

	We consider the problem similarly to a mechanical problem, where all observations generate forces that can be modelled as geometric vectors.
	This kind of problem can be successfully solved with non linear least square. 
	
	Another reason to use non-linear least square is that it is extremely fast, which is one of our requirement for user interaction. 
	So, given an initial road segment network constituted of  \\
	$n$, a set of nodes, \\
	$w$ a set of width relevant to segment (pair of nodes), \\
	$o$ a set of observations \\
	$F_o$ a set of Forces induced by observations and \\
	$F_r$ a set of regularisation forces,\\ 
	we look for the solution of the optimisation problem $S$, so that 
	
	\begin{multline}
		\label{optim.eq.global_optim}
S(n,w) = \argmin_{n,w}(\lVert F_o(o,n,w) + F_r(n,w) \rVert ^2 ) \\
F_o(o,n,w) = F_{kerb}(o,n,w) + F_{object}(o,n,w) + F_{direction}(o,n) \\
F_r(n,w) = F_{position}(n) + F_{length}(n) + F_{width}(w) + F_{angle}(n)  
	\end{multline}	

	Forces generated by observations (detailed in Sec. \vref{optim.method.observation_force}):\\
	$F_{kerb}(o,n,w)$  is generated by the kerb points, \\
	$F_{object}(o,n,w)$ is generated by the street objects surface,\\
	$F_{direction}(o,n)$ is a target road segment angle computed from kerb points. \\
	
	Forces that regulate the results (detailed in Sec. \vref{optim.method.regularisation_force}):\\
	$F_{position}(n)$ limits the nodes toward their initial position, \\
	$F_{length}(n)$ limits the road segment toward their initial length,\\
	$F_{width}(w)$ limits the road width toward their initial width,\\
	$F_{angle}(n)$ limits the angle between adjacent road segment toward their initial values. 
	
\subsection{From raw data to suitable observation and parameters}
\label{optim.method.raw_to_idata}  
	We model the observations as either points or polygons associated with a confidence,
	 a weight and a precision, and a classe of street object.
	 theoretically, our use of polygon is generic enough so any observation type can be modelled (linestrings and points can be buffered), but we also use points for performance reasons.
	The weight allows some distinction for the same object type.
	For instance, an observation of sidewalk which has been observed three times should have an heavier weight than an observation of sidewalk observed once.
	The confidence, (spatial) precision and class of street object are outputs of the detection processes.
	We choose this basic representation of observations to be able to integrate observation coming from various sources, including low level sources.
	
	We use street objects observations as surfaces. 
	Kerb are also streets objects, yet, because they are so influential to determine road surface, 
	and for performance reasons, we model them as points.
	
	We use several wideliy different methods to extract observations of urban features from raw sensing data.
	For some methods we have ot post process the detected urban features so we can properly use them. 
	In particular, we consolidate and fuse kerb and objects detected from mobile mapping lidar (\cite{Serna2014}) because the results are numerous, noisy, and severly overlapping sometime.
	Similarly we fuse pedestrian crossing detection from (\cite{Soheilian2010}) so as to transform individual bands into whole pedestrian crossing.

	Figure \ref{optim.fig.pre_optim} gives an overview of the various data sources we will use in this work and details in what follows.
	\myimageHL{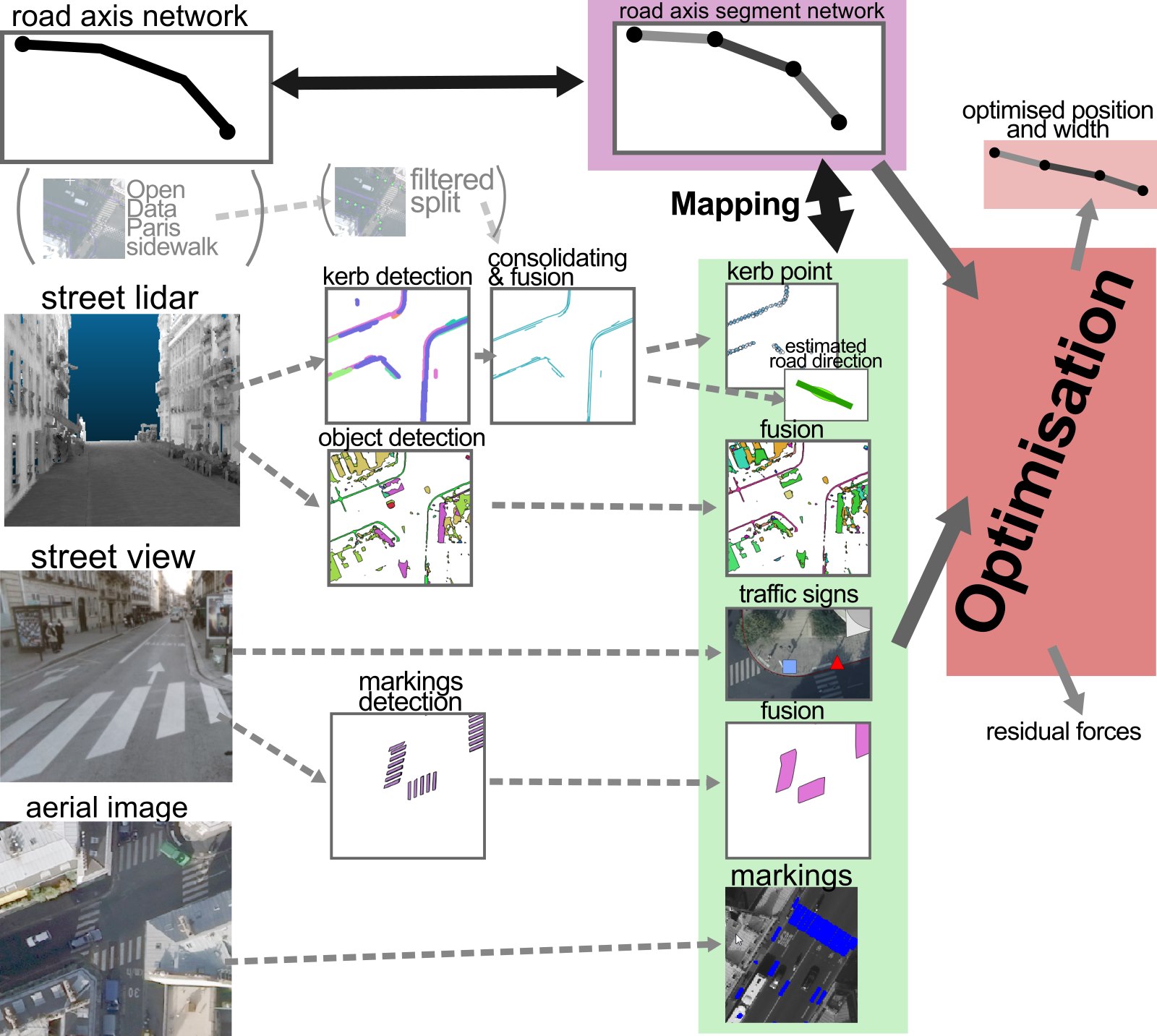}{Data sources for optimisation. The input road axis network is refined into a road axis segment network. We use the detections from several methods. Those detections are processed to become useful observations, which are associated with road segments. The optimisation produces new road segment widths and positions.}{optim.fig.pre_optim}{1}
	
	\subsubsection{Raw data for observation} 
		\label{optim.method.raw_data_to_obs}
		Most of the raw data we use to extract observations come from a mobile mapping vehicle equipped with Lidar and camera (\cite{Paparoditis2012}). 
		It is essential to note that all sensing detections obtained by this vehicle sensors have a geospatial precision limited to the vehicle geo-positioning precision. 
		Although areas with good GPS coverage have precision of under $10 \centi \metre$, the precision is also often in the $40 \centi \metre$ range.
		These data are processed by various methods to extract information.
		\myimageFullPageWidth{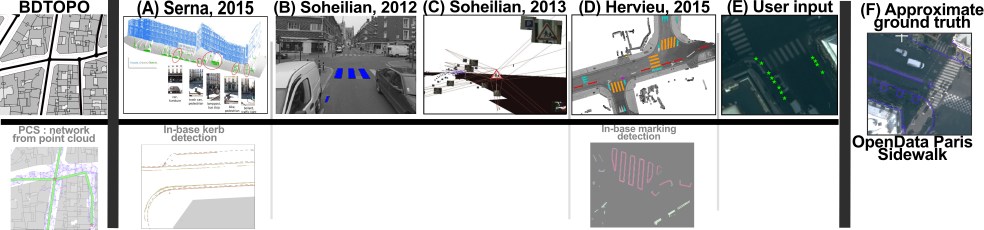}{Overview of observations sources (and potential in-base alternatives). }{optim.fig.data_sources}
		
		\paragraph{(A) :Object from Lidar}
		The main source of data available is the detection of kerbs and other street objects performed on street Lidar.
		We have access to results from \cite{Serna2014}, obtained on datasets from several hundreds of millions of points to 2 billions. 
		(In the future we could also use in-base sidewalk detection from \cite{Cura2015}).
		
		This method performs object detection (including kerb) on point clouds.
		For this, point clouds are rasterized (flattened), then, a gradient of height along with various morphological operation are fed to a random forest learning method.
		Connected components are then extracted, and processed with alpha shape to form polygons.
		We stress that raw results are extremely noisy, especially concerning labelling error for street objects.
		We performed extended filtering and consolidation to transform those detection into relatively reliable observations (We detail that in Section \vref{optim.method.detection_to_observation_for_objects}).
		
		\paragraph{(B) Marking from Street level images}
		We have access to street markings extracted from street level images  (\cite{Soheilian2010}). Those are polygons with basic semantic (type of marking based on width of line, and pedestrian crossing markings), associated with a confidence.
		Not all markings are detected (low recall), but erroneous markings are very uncommon (high precision).
		We performed basic consolidation to aggregate several detection of the same marking found when the vehicle passes several times at the same place.

		\paragraph{(C) Traffic signs from Street level images}
		We also use results from \cite{Soheilian2013a} method, which detects traffic signs based on street level images through a multiview 3D reconstruction.
		Results are semantic points, with confidence and precision.
		Again the result contains very few false detections. 
		Furthermore, it should be noted that in Paris, the signs are on the side-walk in the vast majority of cases.
		The result did not require any processing.
		
		\paragraph{(D) Markings from Lidar and aerial view}
		We have access to preliminary results from \cite{Hervieu2015}.
		They use a sophisticated RJMCMC framework to detect detailed markings with more diverse semantic from aerial images or rasterised Lidar. 
		(In future work we could integrate marking detection from \cite{Cura2015}.)
		 
		 \paragraph{(E) User defined kerb points}
		 Interaction is essential, therefore, we consider another observation raw data which is user defined points or polylines (converted to points (See Sec. \ref{optim.method.lines_to_point})) representing kerbs.
		 These geometries are associated with weight, so that they typically more important than observations from automated sensing. 
		 User defined kerb points are stored in a separate table, to allow easy logging and backup.

		\paragraph{(F) Kerb from GIS open source data}
		Lastly Open Data Paris \footnote{\url{http://opendata.paris.fr/}} provides a side-walk layer, which contains mixed data about sidewalks and urban features, including kerb polylines.
		Because this layer does not contain only kerbs, but also mixed information, using it besides visual ground truth is difficult.
		We filter this data to create an approximated quantitative ground truth, removing parts too small and using data semantics.
		
		Beside creating an approximate ground truth, we can also create almost perfect kerb observations from these sidewalks (we convert again polyline to points).  

	\subsubsection{from lines to points}
	\label{optim.method.lines_to_point}
		Our optimisation method uses kerb observation in the form of points.
		In several cases, we have then to convert polylines to those points.
		(Using Open Data Paris sidewalk, using kerb detection from Lidar, using linestring user input).
		For this we split the lines so that no segment is bigger than  $l_1$ (usually a few metres), then we assign the weight $\frac{SegmentLength}{l_1}$ to each segment node.

	\subsubsection{(A) Street object detections to observations}
		\label{optim.method.detection_to_observation_for_objects}
		Curb detection from Lidar data is extremely noisy. 
		Therefore, we had to resort to sophisticated filtering and consolidation.
		Part of the noise comes from the fact that the sensing vehicle passed several times in the same street, while its geo referencing systems encounter drift of up to $0.4 \metre$.
		We use a two step approach, with first consolidation of data and geometrical filtering, then contextual filtering.
		In both steps, we use data that could be reconstructed from sensing, therefore not loosing any genericity.
		We perform the processing using SQL queries and morphological operation with PostGIS.
		
		Starting from noisy kerb points with confidence, the goal is to consolidate the points into polylines.
		
		Starting from initial detected lines, we dilate (minkowsky sum with a disk) the lines with a radius of $0.4 \metre$ (line spatial precision), then perform a boolean union of all obtained surfaces (unioned surface).
		For each new surface obtained, we transfer the confidence from initial points using weighted mean. 
		We perform a straight skeleton (\cite{Aichholzer1996}) on unioned surface, then filter the straight skeleton resulting lines to remove most of the minor radial segments.
		A simplification step reduces the geometrical complexity of lines (generalisation).
		Further contextual filtering must be performed, where (short) lines too close to buildings and crossing road axis are removed.
		 (See Figure \ref{optim.fig.kerb_detection_correction}).
		
		\myimageHL{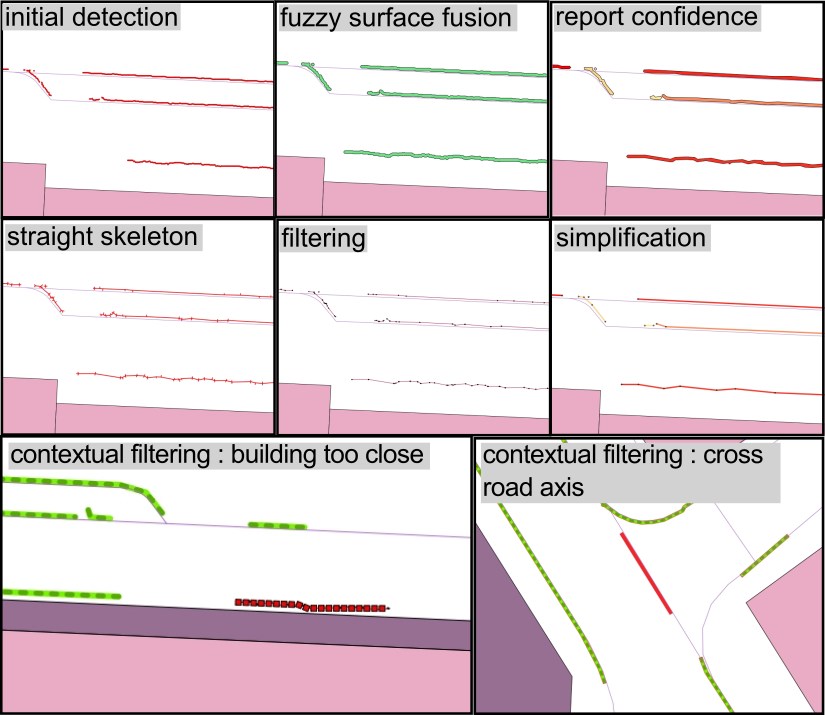}{Input kerb detection is consolidated and filtered (both geometrically and with context) to produce suitable observations.}{optim.fig.kerb_detection_correction}{1}
  
		The cleaning process produces lines as output. 
		We use these lines in two forms, as kerb points, and as kerb line segments.
		The first usage is to create weighted points (Sec. \ref{optim.method.lines_to_point}).

		The second usage is to create segments of (almost) constant length $l_2$ that will be used to robustly determin a target slope of the road segment.
		
		Please note that the example Figure \ref{optim.fig.kerb_detection_correction} is a very favorable detection case. 
		(See Figure \vref{optim.fig.kerb_detection_correction_limit} for more common and challenging data.)
		
	\subsubsection{Road axis network}
		\paragraph{Raw data for road axis network}
		We use road axis geometry with approximate road width from IGN BDTopo.
		They consist of polylines with attributes.
		We create a PostGIS Topology from these axes, that is a graph of road axis polylines, with polylines being the edges of the graph, and the nodes being the intersections. 
		Road axis precision is generally metric\footnote{\url{http://professionnels.ign.fr/bdtopo}}, approximate width precision is usually several meters. 
		
		\paragraph{Preparing network for optimisation}
		The optimisation works on segments, therefore, we have to convert polylines into segments, while keeping its filiation to the road axis network topology.
		We perform these operations in base with custom SQL queries.
		Retaining the filiation allows to keep track of each road segment approximate width,
		as well as futur usage and analysis of results.
		
	\subsubsection{Linking observations to street segments}
	 
		By design, each observation affects at most one street segment.
		Each observation must be attributed to a road segment, which we will call "match". We do not use "mapping" whihch is more accurate in database context to avoid any confusion with geographical mapping.
		Finding the optimal match between observations and road segments is a combinatorial problem.
		The chosen optimisation framework (robust non-linear least square) is not powerful enough to optimise simultaneously on matching and road model parameters.
		
		However, we rely on a simple matching method (closest road surface), because streets tend to be less wide than city blocks, and because initial road axis and road width provide a rough road surface (precision of a few meters).  
		Outside of intersections, these properties indicate that an observation can reasonably be matched to only one street, the other one being too far away to be considered anyway (See Fig. \vref{optim.fig.matching}). 		\\
		 
		\myimageHL{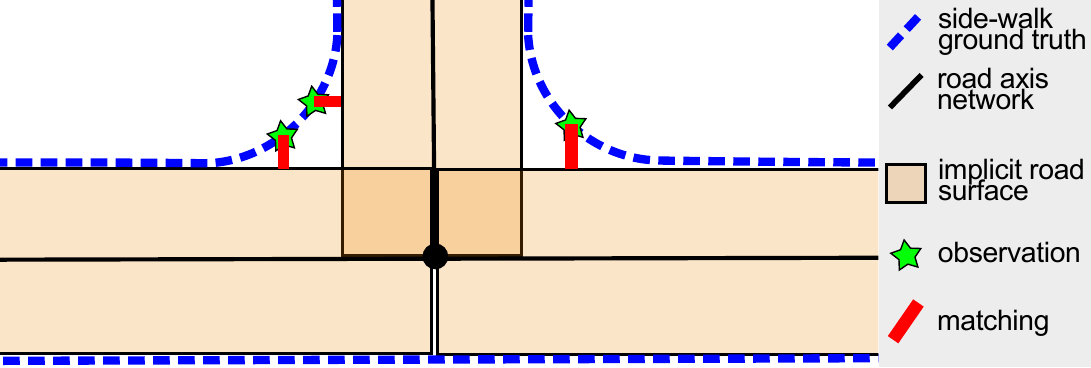}{In intersections, kerb observations can not be used as the road model is not adapted. Using these observations would lead to incorrect results. }{optim.fig.matching_in_intersection}{1}
		
		A notable exception is for observations near intersections, where our road model is clearly not adapted to curved kerb and curbstone inside the intersection (See Fig. \ref{optim.fig.matching_in_intersection}).
		Please note that these observations are also amongst the noisiest due to limited curbstone height in Paris intersections.
		Therefore, we remove observations close to intersection surfaces. 
		To determin intersection surface, we use Streetgen (\cite{Cura2015a}).
		(Still, those observations could be used to optimise the turning radius in StreetGen road model.)

		This matching process (assigning observations to the closest road surface) is performed in base with carefully written queries.
		In theory, matching would require to consider $N_{observation}*N_{road segment}$ possibilities, which would become intractable at a city scale ($500k*50k$ for Paris).
		However, we use database acceleration structures (Rectangle tree: GIST) so that only observation-segment pairs spatially close enough are considered.
		Processing is only a few seconds for about half an arrondissement (Paris is divided into 20 arrondissements), and a few minutes for the whole city of Paris.
	
	\subsubsection{Estimated road direction of road segment from observations}
		\label{optim.method.direction_from_observation}
		When matches between kerb observations and road segments have been computed, we use kerb observations to estimate a road direction (see Fig \ref{optim.fig.slope}) for each road segment having observations.
		All observation segments are weighted by their length, then the direction of each road segment is computed and a weighted median performed on these directions. 
		We choose the weighted median because it is not very sensitive to observation noise, while being fast to compute.
		The estimated road direction is then the weighted mean of the direction of observations having a direction close to the weighted median direction (by a threshold we experimentaly fix to 20 \degree ).
		
		For each kerb observation segment \\
		split kerb observation linestrings to weighted kerb observation road segments (weight = length) \\
		compute direction $d_i$ for each weighted road segment $(s_i,w_i), i \in [1,N]$ \\
		compute directions weighted median \\$d_{wmedian}= weighted\_median(\lbrace (d_i, w_i), i \in [1,N]\rbrace)$\\
		estimated road direction =\\$weighted\_mean(\lbrace (d_i, w_i)\rbrace, i \in [1,N],\left| d_i- d_{wmedian} \right| < th ) $ 
		Where $th$ is a threshold we experimentaly fix to 20 \degree.
	
		\myimageHL{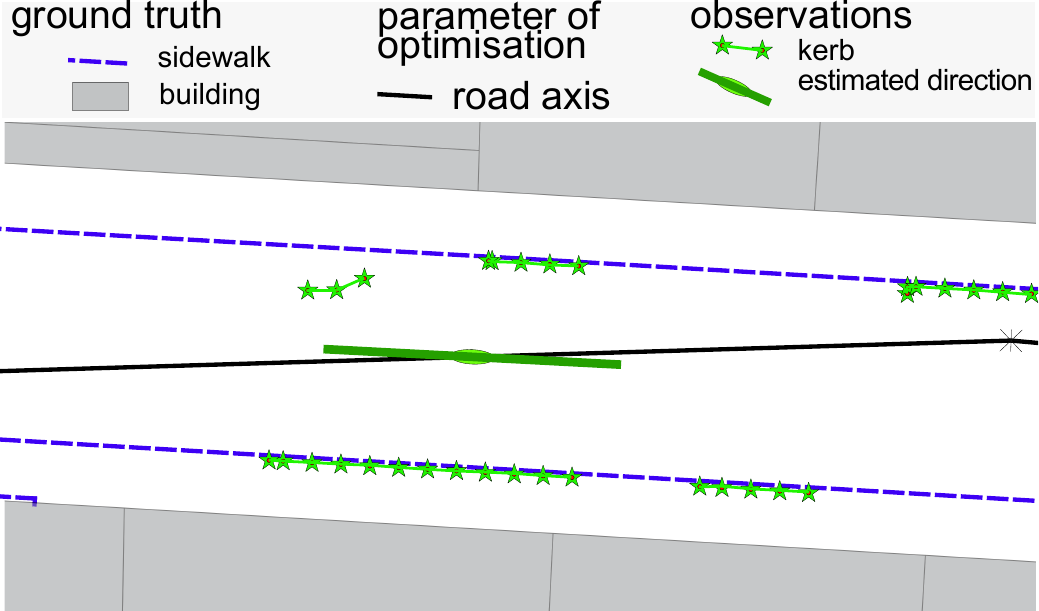}{Estimated road axis segment slope based on weighted median of kerb observations. The estimated slope is coherent with the ground truth, and is very different from the road axis to be optimised.}{optim.fig.slope}{1}

\subsection{Observation and regularisation forces}
	We model the optimisation problem as a mechanical problem, where each observation generate forces on the road model variables, and additional regularisation forces limits the variations of variables.
	
	Processing of Section \ref{optim.method.raw_to_idata} produces a set of connected road segments with an approximate width (parameters to be optimized), and observations (weighted points or polygons, with confidence) matched to those road segments.
	
 	For our optimisation framework (Ceres-Solver), all forces are called constraints, with force values being called residuals, and force directions being called Jacobian (for each parameters).
 	
 	Therefore each force defines a residual (whose squared value will be minimised), and a direction, that will be used to solve the optimisation problem.
 	
	 \subsubsection{Observation forces ($F_o$)} 
	 \label{optim.method.observation_force} 
	
	 Observations generate forces over node position and road segment width (See Fig. \ref{optim.fig.forces}). 
	
	 \myimageHL{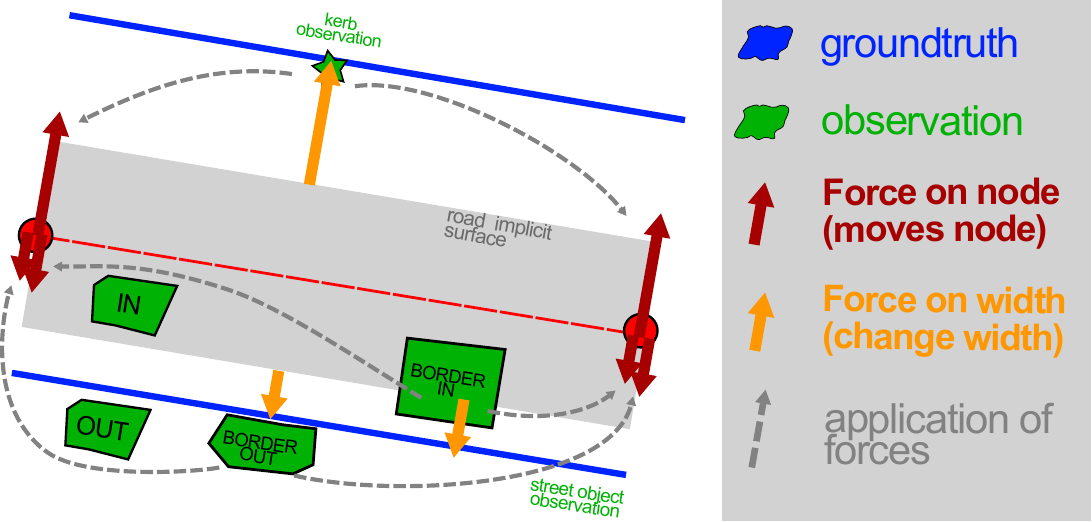}{Example of forces induced by observations.}{optim.fig.forces}{1}
	
	We define three forces from observations, as seen in equation \vref{optim.eq.global_optim}\\
	 $F_o(o,n,w) = F_{kerb}(o,n,w) + F_{object}(o,n,w) + F_{direction}(o,n)$
	 
	 The idea is always that an observation generates forces that tend to change the road model parameters so the road model fits this observation.
	 
	 Each force from observations exists in two versions; one affects node position, the other affects edge width.
	 The residuals are identical, but the Jacobian changes.
		
		\paragraph{Force from kerb points ($F_{kerb}$)}
			Each kerb point observation generates forces so the road surface border passes on this kerb point (See Fig. \vref{optim.fig.forces}). Both road segment nodes and road width are affected. 
			We use the same notations as in the open source implementation\footnote{\url{https://github.com/Remi-C/Network_snapping/blob/master/using_ceres/Constraints.h\#L409}}.)
			Given a road segment $N_iN_j$ of width $w$, a kerb observation point $Ob$.
			The goal is to find the orthogonal distance between $Ob$ and implicit road surface (i.e. implicit rectangle formed by road segment $N_iN_j$ using road width $w$).\\
			We compute $\overrightarrow{N_p}$ the normal of the plan $P$ containing $Ob,N_i,N_j$ with \\
			$\overrightarrow{N_p} = \overrightarrow{ObN_i} \times \overrightarrow{N_iN_j}$.\\
			Then $d = \frac{\lvert\overrightarrow{N_p} \rvert}{\lvert\overrightarrow{N_iN_j}\rvert } -\frac{w}{2}$.

			$d$ is the residual (whose squared value will be minimized).

			For the version of the force affecting node positions,
			$N_i$ and $N_j$ changes happen in $P$, in an orthogonal direction to $\overrightarrow{N_iN_j}$, whose director vector is computed with \\
			$V_{ja} = \frac{\overrightarrow{N_iN_j}}{\lvert\overrightarrow{N_iN_j} \rvert} \times  \frac{\overrightarrow{N_p}}{\lvert\overrightarrow{N_p} \rvert}$.
			
			For the version of the force affecting width, the proposed width $w_n$ is so $d=0$, that is \\
			$\frac{w_n}{2} = \frac{\lvert\overrightarrow{N_p} \rvert}{\lvert\overrightarrow{N_iN_j}\rvert } $.
			
		\paragraph{Object observation ($F_{object}$)}
			We define another observation forces based on surface-like object observations (See Fig. \vref{optim.fig.forces}).
			Objects can be arbitrary polygons potentially having several inner holes.
			
			The idea is similar to the kerb point force: the force is based on the distance between objects and implicit road surface border.
			As opposite to point to segment distance computation of kerb observations, the necessary segment to polygon distance for objects is not easily done.
			For this, we use GEOS \footnote{\url{https://trac.osgeo.org/geos/}} to find the distance between the implicit road surface (road segment and road width) and the object.
			
			Each object class has an expected behaviour regarding road surface (IN, OUT, BORDER IN, BORDER OUT), that is used in the force.
			This distance also takes into account the expected distance between a class and the border of the road if the class is of type "BORDER".
			That is, objects of this class are expected to be at a given distance of the border.
			For instance, in Paris, a barrier is expected to be $0.2 \metre$ from the border of the road (kerb).
			
			Unlike points, a special case occurs when an object in neither entirely in or entirely out of the implicit road surface.
			In this case, we consider that the distance is proportional to the percentage of the object surface that is inside the implicit road surface, i.e.\\
			 $\frac{Area(Object \cap RoadSurface)}{Area(Object)}$.
			This definition is generic enough to work well with any object we may use, and makes sense at the same time when considering the surface as the probabilistic location the object may be.
			
			The version of the force that affects nodes has similar direction: orthogonal to edge axis in X,Y plan.
			The version of the force that affects width simply propose a new width so d=0 .
		
		\paragraph{Road segment azimuth from kerb points ($F_{direction}$)}
			
			For the two previous types of forces, the observations are considered individually.
			However this may lead to an ill conditioned problem.
			Strongly unbalanced observation density may stick the optimization in a local minimum.
			Intuitively, two points (at least) are necessary to determine a segment direction.
			To solve this problem, we use all observations affecting the road segment to determine a probable road segment direction $d_o$ (See Sec. \vref{optim.method.direction_from_observation}).
			
			The force is defined as follows.
			Given a probable target road segment direction estimated from kerb observation $dir_o$, the force rotates the road segment around its centre point so the road segment direction $dir_s = dir_o$.
			
	\subsubsection{Regularisation forces ($F_r$)}
		\label{optim.method.regularisation_force}
		Observations generate forces over node position and road segment width (See Fig. \ref{optim.fig.forces}). 
		We define four regularisation forces, as seen in equation \vref{optim.eq.global_optim}\\
		$F_r(n,w) = F_{position}(n) + F_{length}(n) + F_{width}(w) + F_{angle}(n)$
			
		These regularisation forces keep the optimisation result close to the optimisation initialisation (See Fig. \ref{optim.fig.forces_conservative}). 
		
		\myimageHL{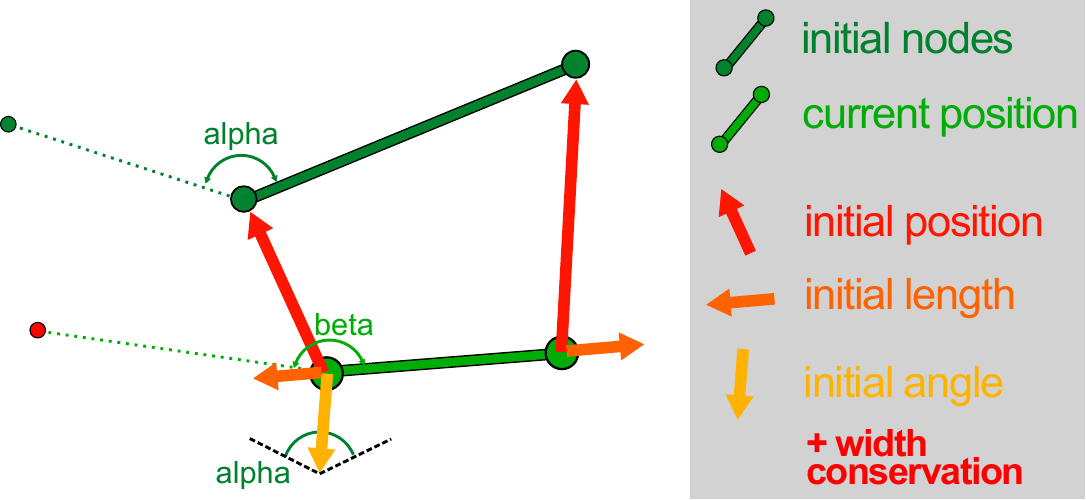}{Regularisation forces that work to preserve initial values of parameters. The "initial position" force tends to preserve each node initial position. The "initial length" force tends to preserve each road segment length. The "initial angle" force tends to preserve initial angle between road segments. The "initial width" tends to preserve the initial width of each road segments.}{optim.fig.forces_conservative}{1}

		In other words, the regularisation forces works to push the parameters towards their initial values, so as to limit changes and avoid large variation of parameters.
		Such regularisation are needed for three reasons.
		First, the initial road segment and road width (optimisation initialisation) are not far from the optimal solution (few meters).
		Second, road segments might not have any associated observations at all. As such,  they could potentially be moved hundreds of meters.
		Third the optimisation does not perform validity checks on optimised road segments (for instance, we do not check if a new road segment position makes it intersect another segment, which would be forbidden in a topology).
		
		The only parameters of the optimisation are road node position and road segment width, thus, in theory, only two regularisation forces would be necessary.
		Yet, regularisation forces are meant to preserve different properties of the initialisation.
		Lets take the example of a perfect road network, with perfect with and node positions values, but that is translated 1 metre North.
		Conserving the initial node position would prevent from translating the nodes 1 metre South to get the correct result.
		However, stating that angles between road segments shall be preserved allows the freedom to perform the translation, but still preserve the overall road network organisation.

		We describe each of these regularisation forces (\\ $F_{position}, F_{length}, F_{width}, F_{angle}$) in the following paragraphs.
		
		\paragraph{Distance to initial road segment position ($F_{position}$)}
			Before the optimisation starts, the initial road segment node positions (X,Y,Z) are stored.
			Then the resisting force is defined for each node as the euclidean distance between the initial position and its current position.
			The Force direction is $\overrightarrow{N_ioldN_i}$.
			
		\paragraph{Distance to initial road segment length ($F_{length}$)}
			This force is necessary to limit the deformation of the road segment network.
			Before the optimisation starts, the initial road segments length are stored.
			The resisting force is defined as trying to maintain the initial road segment length.
			Force intensity is defined by $ F_i = Length_{initial} - Length_{current}$,
			the force direction is $\overrightarrow{N_iN_j}$, with direction depending on the sign of $F_i$.
			
		\paragraph{Distance to initial road segment width ($F_{width}$)}
			This force is particularly useful to limit large width changes when only a few noisy observations are present.
			The initial width is stored for each road segment. 
			Then the resisting force is $ \lvert InitialWidth-CurrentWidth \rvert$.
			This force only change width, and has no effects on node positions.
			
		\paragraph{Distance to initial pair of road segment angle ($F_{angle}$)}
			For each successive road segment node inside a road axis, we store the angle  $\widehat{N_iN_jN_k}$.
			That is, when possible, we associate to a node $N_j$ its initial angle ($\hat{A_j}$) with previous ($N_i$) and next ($N_k$) node.
			
			The force direction is then the bisector of this this angle $\overrightarrow{B_j}$, which is computed with\\
			$\overrightarrow{B_j} =  \frac{\overrightarrow{N_jN_i}}{2 * \lvert\overrightarrow{N_jN_i} \rvert} + \frac{\overrightarrow{N_jN_k}}{2 * \lvert\overrightarrow{N_jN_k} \rvert} $
		
			Given the current positions of the 3 nodes, we look for the new position of $N_j'$ so that  \\
			$\widehat{N_iN_j'N_k}$ = $\hat{B_j}$.
			Finding the actual distance $d$ to the new position of the node $N_j'$ along the bissect is complicated,
			 as it require to solve a system so that \\
			 $(\overrightarrow{N_jN_i}-d*\overrightarrow{B_j})\times (\overrightarrow{N_jN_k}-d*\overrightarrow{B_j})\cdot \overrightarrow{N_p}= tan(\hat{A_j}) * (\overrightarrow{N_jN_i}-d*\overrightarrow{B_j})\cdot (\overrightarrow{N_jN_k}-d*\overrightarrow{B_j})$
			 \\where \\
			 $\overrightarrow{N_p} = \frac{\overrightarrow{N_jN_i} \times \overrightarrow{N_jN_k}}{\lvert \overrightarrow{N_jN_i} \times \overrightarrow{N_jN_k} \rvert} $.
			 
			 Instead, we approximate computing of $d$ by considering that $\lvert \overrightarrow{N_jN_i} \rvert = \lvert \overrightarrow{N_jN_k} \rvert $.

\subsection{Optimisation}
	As seen in equation \vref{optim.eq.global_optim}, the optimisation problem is formed of optimisation variables (road segment node position, road segment width), and forces (from observations, for regularisation).
	
	In this section we consider the optimisation process in itself, from how we influence it with several meta-parameters (weight, bounds, loss function), to which optimisation strategy we use, to how we can use the resulting road model to generate a street model with StreetGen. 
	
	\subsubsection{Meta parameters}
		First we use several level of weights to balance the level of confidence on different data and prior knowledge, second we bound the variation of the road model variable, then we weather the influence of outliers using loss functions.
		\paragraph{Weights}
		
		\myimageHL{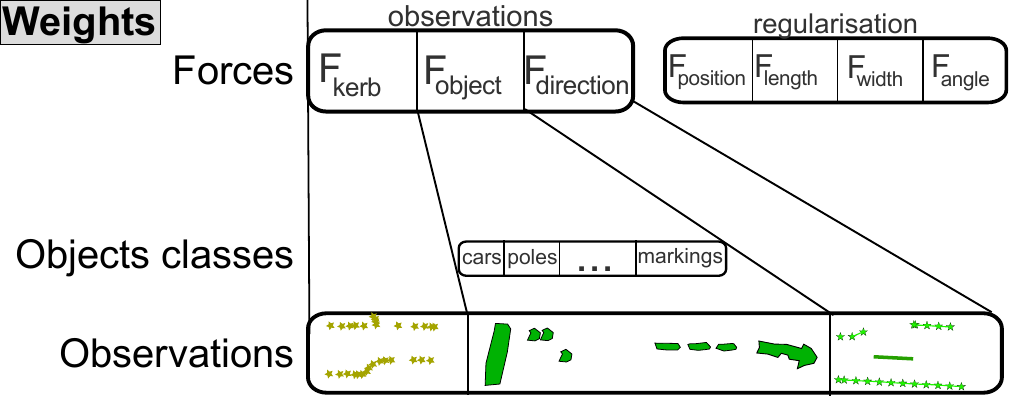}{Several levels of weight allow to set the level of confidence in various data and prior knowledge.}{optim.fig:weight}{1} 
		We use several levels of weights to adapt the optimisation process to data sources (see fig \ref{optim.fig:weight}).
		 
		At the lower level, each observation is weighted. 
		It controls how much faith the user has in this particular observation.
		If the observation is an object, each class of object is also weighted.
		This is necessary because some street objects are much harder to detect than others (for instance poles are easier to detect than trash cans), then this weight allows to express a preference between object types.
		
		At the higher level, each type of force has a generic weight.
		This is crucial to balance between the confidence in observations and the confidence in the initial solution,
		or between forces.
		For instance, a user may know that the input road axis network is very precisely positioned, but with imprecise road width.
		This force-level weight is also what enables use of our methods in different scenarios.
		For instance, if observations are from noisy sensing, the weight of regularisation forces will be higher.
		On the opposite, if observations are from legacy GIS data, the regularisation force weight will be low, because we grant much more confidence to observations.
		
		Choosing all these settings then depends on knowledge about input data and sensing data.
		We used an experimental approach to choose these settings.
	
		\paragraph{Bounds}
		Bound are used to limit the optimisation search space.
		For instance the road width is not realistically expected to vary more than by a dozen meters, neither a node to move more than a few meters.
		Bounds both limit the absolute range (e.g. width should be between 1 and 20 meters), and the relative range of parameters (e.g. width variation should be less than 10 meters).
	
		\paragraph{Loss functions}
		Least square frameworks are very sensitive to outliers.
		Intuitively, an absurdly high value would have a very large squared value,
		which would in turn dominates the other regular values.
		Ceres-solver allows to use a classical solution to this problem: loss functions.
		Basically, instead of optimizing on $x^2$, we optimise on $f(x)^2$, where f is a function that acts like the square function at low scale, but is much flatter at high scale.
		In our case, we choose the "Soft L1" $f(s) = 2*(\sqrt{1+s}-1)$.

	\subsubsection{Meta strategy}
	We tested two strategies. The first is to optimise all parameters at the same time.
	The second strategy is to successively optimise for width and position, until no more improvement is reached.
	The first strategy is canonical and produces the best results, at the price of slightly more lengthy computation.
	The Second strategy is faster, but can be stuck in local minima and can produce worst results.
	As such, we consider the second strategy is not worth to be used, and favour the first strategy.
	
\subsubsection{Generating streets from optimised road model}
\label{optim.method.optim_sg}
	We use StreetGen to generate streets from the optimised road model (road axis segment network and road segment width), as depicted in Figure \ref{optim.fig.sg_optim}.
	
	\paragraph{regrouping road segments}
	\myimageHL{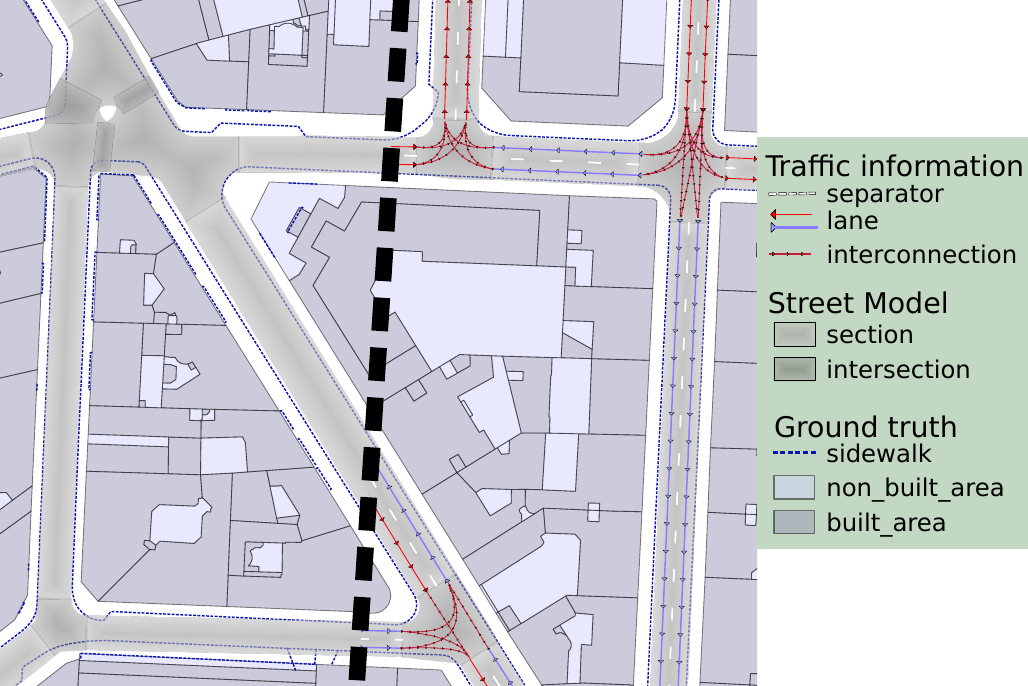}{We use the optimised road model (kerb observation from sensing) as input to StreetGen to generate a complete street model. Note that turning radius have not been optimised.}{optim.fig.sg_optim}{1}
	
	It is important to note that the StreetGen street model is based on polyline road axis associated with one width per polyline,
	while the optimisation road model is based on segment of road axis associated with one width per segment.
	
	Before optimisation, each polyline is broken into segments.
	
	A simple solution to use streetgen on the optimisation result would then be to consider that each road segment is in fact an individual road axis (a polyline composed of only one segment).
	This works in theory, but introduces many useless intersections.
	Indeed in StreetGen intersections are meant to deal with changes of road width and to deal with intersections of more than two road axis.
	In the optimised results, many road segments of the same original polyline may have approximately the same width, and thus should be regrouped.
	
	However, two factors complicate the regrouping.
	First, some segments may not have associated observations, and thus their width has not been optimised.
	Therefore, their width is insignificant and should be harmonized with the width of nearby segments from the same polyline that have observations (if any). 
	Second, only successive segments should be regrouped.
	
	We solve these issue in three steps.
	First, we work on segments having observations and so a significant width.
	For each polyline, we regroup the segments having approximately the same width using the DBSCAN method \citep{Ester1996}.
	The new width is the median width weighted by the number of observation of each segment.
	Second, we work on segments not having observations ("no obs"). 
	We try to find other segments with a significant width in the same polyline.
	If any is found, the segment closest to the "no obs" (and with the most observations is two segments are at the same distance) shares its width with the "no obs".
	If none is found, the original width of "no obs" (un-optimised) is used.
	
	In a third step, in each polyline, we regroup the successive segments with the same width.
	
	\paragraph{topology considerations}
	Our optimisation method does not guarantee to preserve topology (no edge should intersect except in intersection node).
	Instead we use constraints so result is close to initial values.
	It suffices in the vast majority of cases but not all (See Fig. \vref{optim.fig.error_topo}).
	A common error case is when a very small edge is close to an intersection, and therefore has very few chances to have observations matched to him. In this case, this small edge may be moved and end upintersecting another edge, which is a topology error.
	Luckily such topology error detection is easy and fast using ad hoc SQL query exploiting spatial index.

	

 \section{ Results}
	 \label{optim:result} 
	 	 \setcounter{subsection}{-1}
	 In this section, we first consider experiment settings and how to display results and forces (as visual control is as important as qualitative control for a road network).
	 We present results for the pre-optimisation task, such as the transformation of raw data into observations, and the matching process between observations and road segments.
	 Last, we present the optimisation results at different scales with different input data in part \ref{optim.result.optim_result}.
	 
	\subsection{Resources}
		We use Ceres-solver\footnote{\url{http://ceres-solver.org/}} 1.10, the optimisation framework from \cite{Agarwal2016}.
		Our prototype implementation is available as free and open source software\footnote{\url{https://github.com/Remi-C/Network_snapping}}.
		We use a Ubuntu 12.04 OS in a virtual box with dedicated 6 GB of RAM and 6 2.4 GHz Intel Xeon CPU threads. 
		Timings are measured using the standard Cpp library.
	 
	\subsection{Results and Forces visualisation} 
		Optimisation methods are notoriously difficult to control and parametrise,
		especially when input data contain outliers.
		Therefore, we consider essential to have a way to represent the forces, as well as the parameters,
		not only before the optimisation starts, but during the whole optimisation process (each iteration).
		Rather than create a new interface from scratch, we prefer to re-use a well known open source interface: QGIS.
		For each iteration of the optimiser, we compute forces and export them in Well Known Text (WKT), along with the iteration number and a (fake) timestamp, into a comma separated value text file.
		This text file is imported in QGIS and used with the TimeManager extension\footnote{\url{https://plugins.qgis.org/plugins/timemanager}}.
		This extension creates a time slider to slide through the iteration of the optimiser,
		allowing to display both forces, residuals, and parameters. 
		
	\subsection{From raw data to Observation} 
		In Section \vref{optim.method.raw_data_to_obs}, we introduced pre-processing methods to filter and consolidate urban feature detections into suitable observations.
		The pre-process is especially necessary for kerb dection from Lidar data (Section \ref{optim.method.detection_to_observation_for_objects}).
		
		Overall, the filtering and consolidating of kerb detection from Lidar data produce a much more tractable result than the initial one.
		The process is a few minutes long for about half a million detections (mostly due to the Straight Skeleton).
		The process can not remove all the errors, in particular because confidence measures are not giving much information.
		However, when observation density is sufficient, the optimisation seems to be robust enough to deal with the remaining noise.
		We compiled small examples of the remaining errors in Fig. \ref{optim.fig.kerb_detection_correction_limit}. 
		
		\myimageHL{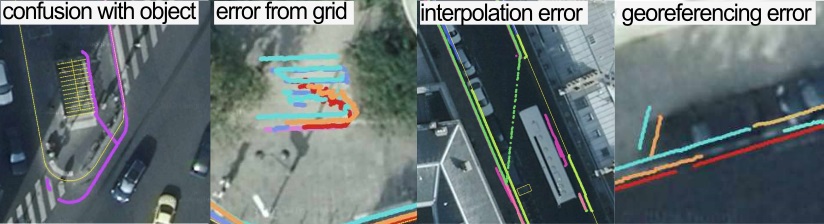}{Input kerb detection is consolidated and filtered (both geometrically and with context) to produce more reliable observations.}{optim.fig.kerb_detection_correction_limit}{1}

	\subsection{Observations matching}
		Overall, the straightforward matching approach (closest implicit road segment surface) seems to work very well for kerb detection, and is fast thanks to our indexed approach (few seconds for sensing area).
		Object observation matching also seems to be correct (Fig. \ref{optim.fig.matching}).
		We could find one case with incorrect matching, unsurprisingly in an intersection area (Fig. \ref{optim.fig.matching}, right illustration, red circle).
		
		\myimage{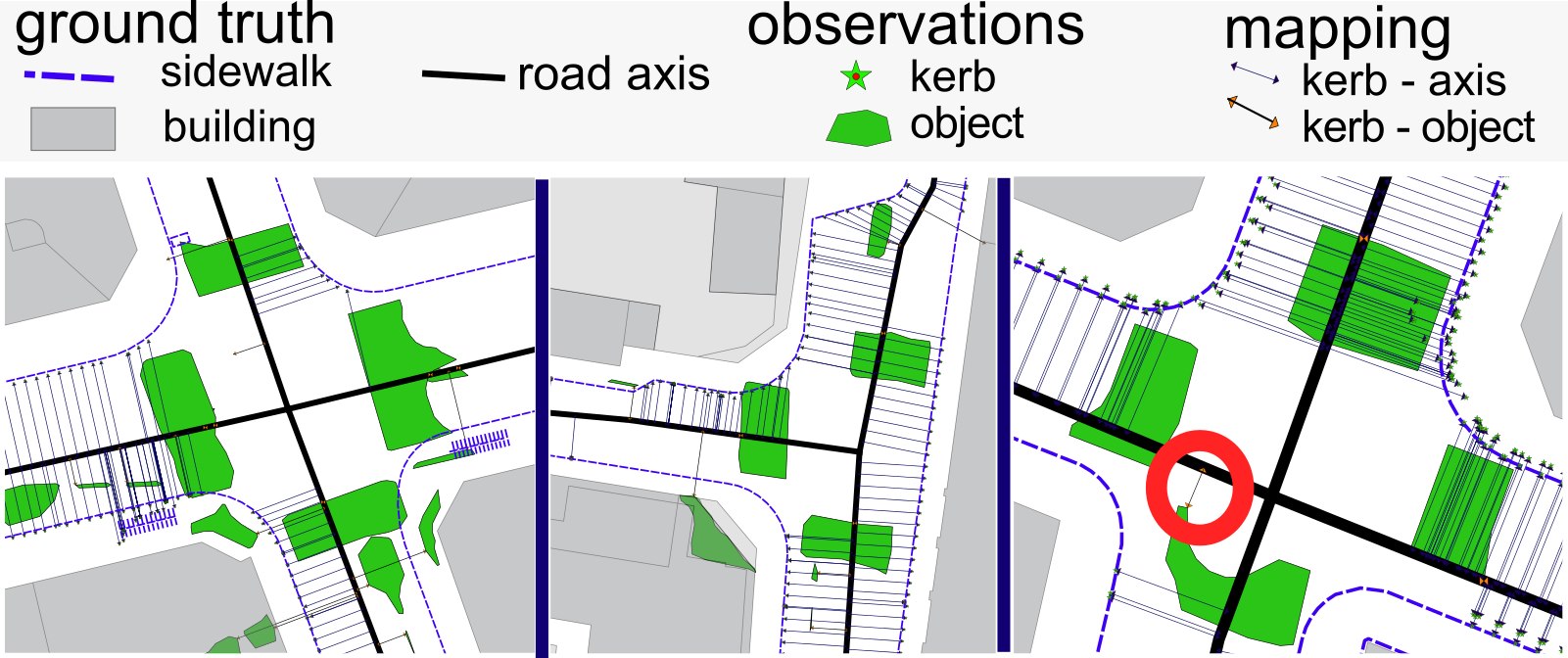}{Observations are attributed to the closest implicit road surface, computed using road axis and road width. This simple matching works well, except inside intersections (Third figure, red circle).}{optim.fig.matching}
	
	\subsection{Optimisation results}
		\label{optim.result.optim_result}
		\subsubsection{Result evaluation}
		
		Multiple factors introduce errors in the optimisation process, which complicates result evaluation (See Fig. \ref{optim.fig.errors_sources}). 
		\begin{itemize}[noitemsep,topsep=0pt,parsep=0pt,partopsep=0pt]
		\item The ground truth is not perfect (it has not been updated since 2010) and it mixes side-walk with other urban features.
		\item The observations are noisy and sometime sparse.
		\item The input road axis network is not split enough. As our optimisation method does not change topology, this leads to an under-parametrised problem (or over constrained).
		A least square problem is by nature mathematically over-constrained (observations are redundant and contradictory).
		However, we refer to the fact that a human performing this optimisation would change the number of parameters by merging edge segment or splitting them.
		\item The chosen road model is not powerful enough to model all types of roads.
		\end{itemize}
		
		We design various strategies to limit these errors, allowing to analyse the influence of each source of error independently. 
		\myimageFullPageWidth{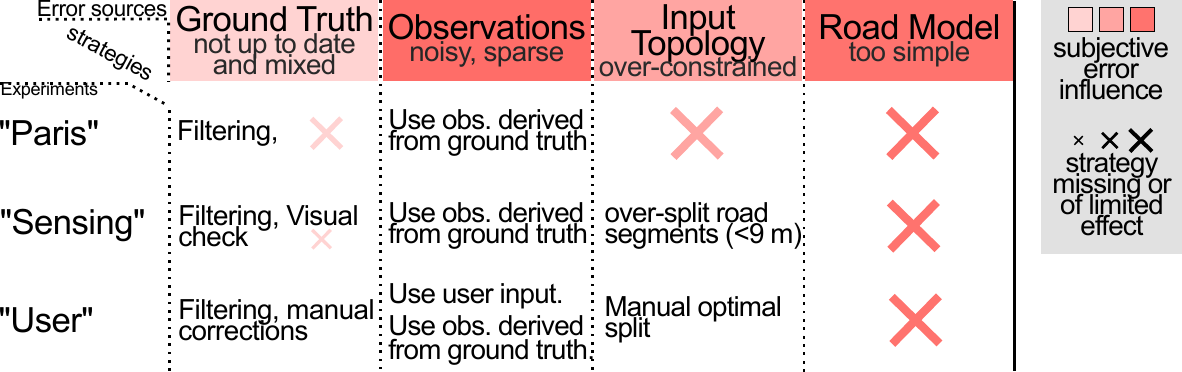}{Multiple factors introduce errors in the optimisation process, which complicates result evaluation. We design various strategies to limit these errors, allowing to analyse the influence or each source of error independently.}{optim.fig.errors_sources}
		 
		\paragraph{Perfect observation derived from ground truth}
		Optimisation results \\
		strongly depend on observations quality (sparsity and noise).
		To remove this bias from the analysis, 
		some optimisation are not performed on actual observations,
		but directly on ground truth side-walk points from Open Data Paris.
		
		\paragraph{User input to complement observation sparsity} 
		Lack of observations has a big impact on optimisation.
		Therefore, we generate the best possible results with our noisy and sparse observations from sensing (Sec. \ref{optim.method.raw_data_to_obs}).
		In a second step, we introduce user inputs that will be used in the optimisation.
		Those user inputs are similar to StreetGen : user defined curbstone (kerb) points.
	
		\paragraph{Qualitative evaluation by measuring residual distance to sidewalk}
		We design an approximate qualitative measuring.
		The aim is to measure how much ground truth road surface is covered by the optimised implicit road surface formed by road axis and road with.
		There is no road surface ground truth for Paris, therefore, we use OpenDataParis side-walk border layer to get Paris sidewalks.
		We regularly place points on the ground truth kerb, avoiding the intersection area.
		The distance between kerb point and closest implicit road surface is then computed.
		The distance will be very small if implicit road surface fits well with the side-walk, and large otherwise.
		This measure can not be perfect because the side-walk layer also contains some street objects (not a perfect ground truth).

		\subsubsection{Experiment areas}
		 \label{optim.result.experiment_areas}
		\paragraph{Choosing experiment areas}
			Three different sized areas are chosen, each having a specific interest (see Fig. \ref{optim.fig.area_of_experiments}).
			\myimageHL{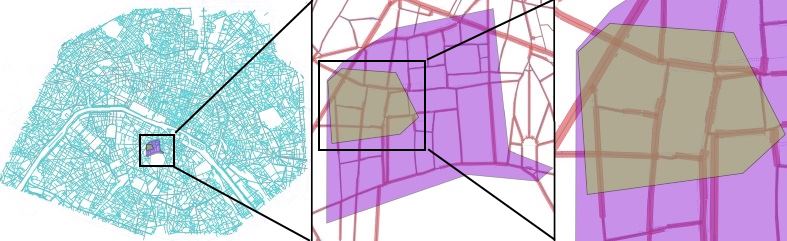}{Three different experiment areas for various experiments, from the whole Paris to half an arrondissement to a few streets.}{optim.fig.area_of_experiments}{1}
			
			\begin{itemize}[noitemsep,topsep=0pt,parsep=0pt,partopsep=0pt]
			 \item The first area ("Paris") corresponds to the whole city of Paris.
			Sensing data is not available for all of Paris.
			Therefore, we do not use observations from sensing data, but instead, we create observations from ground truth (Open Data Paris Sidewalk).
			This area is used to demonstrate scalability, analyse the use of constraints, and check the optimisation robustness in extremely complex parts of the road axis network.  
			
			\item The second area ("Sensing") is the whole area where sensing data is available.
			Road types and observations are very diverse.
			This area is used to determine how useful a fully automatic optimisation process would be,
			as well as evaluate computing time for this amount of object observation.
			
			\item The third area ("User") is a small area with sensing data available.
			We purposefully chose the most challenging area regarding road surface complexity.
			Due to its small size, road axes can be manually split into the relevant number of road segments, thus eliminating the over-constrained bias of the evaluation.
			A thorough visual control is also possible.
			Last, the moderate size allows some experiments on the usefulness of user inputs. 
			\end{itemize}

		\paragraph{Numerical facts about experiment areas} 
			Table \ref{optim.tab.experiment_area} shows an overview of experiment areas size and content.
			Optimisation computing time is between a few minutes ("Paris"), a few seconds ("sensing") and less than a second ("user") when no object observation is used. 
			Using object observations greatly slows the optimisation process as it relies on the GEOS library to perform geometrical computing at a great cost.
			As such, optimising time is a few minutes for "sensing" and a dozen of seconds for "user".

			\begin{table*}[]
				\centering 
				\scriptsize 
				\begin{tabular}{ccccccccc}
				\shortstack{Area\\ \space} & \shortstack{\# edges\\ \space} & \shortstack{\# nodes\\ \space} & \shortstack{\# Groundtruth \\ Observation}&\shortstack{ \# Kerb\\ Obs.} &\shortstack{ \# Car \\ Obs.}& \shortstack{ \# Markings\\ Obs.} & \shortstack{\# all objects\\ Obs.} & \shortstack{\# User \\ Input}\\  
				Paris &38.8 \kilo &45 \kilo &522 \kilo & & & & & \\
				Sensing &186 &215 &2343 &11.7 \kilo &390 &1333 &5523 & \\ 
				User &62 &64 &475 &2628 &145 &136 &2011 & 12 places\\ 
				\end{tabular} 
				
				\caption{\label{optim.tab.experiment_area} Facts for ground truth, parameters and observation for the three experiment areas.}  
			\end{table*}
			 
		\subsubsection{Results on "Paris" area}
		The first subjective result is that optimisation is very successful, the fitting of the model is greatly increased.
		The qualitative result (See Table \ref{optim.tab.whole_paris}) confirms this (median distance to ground truth diminished by a factor of ten, from $1.547 \metre$ to $0.104 \metre$).
		At the same time, because optimisation is performed on ground truth observations, the fact that the result is not perfect shows that the model is not expressive enough for all types of roads, and that it is over-constrained.\\
		However, these two bias (model too simple, over-constrained) do not prevent the results to be very close to ground truth.

		The second result on the "Paris" area is that optimisation scales very nicely.
		The area represents about $170 \kilo$ parameters, and around half a million observations, plus regularisation constraints. 
		The entire optimisation is done in 3 to 4 minutes, with about 30 seconds spent on reading data and constructing the problem (in a Ceres-solver meaning).
 
		The third result is that optimisation is robust, even when the input road network is extremely complex (see Fig. \ref{optim.fig.example_complex}).
		\myimageHL{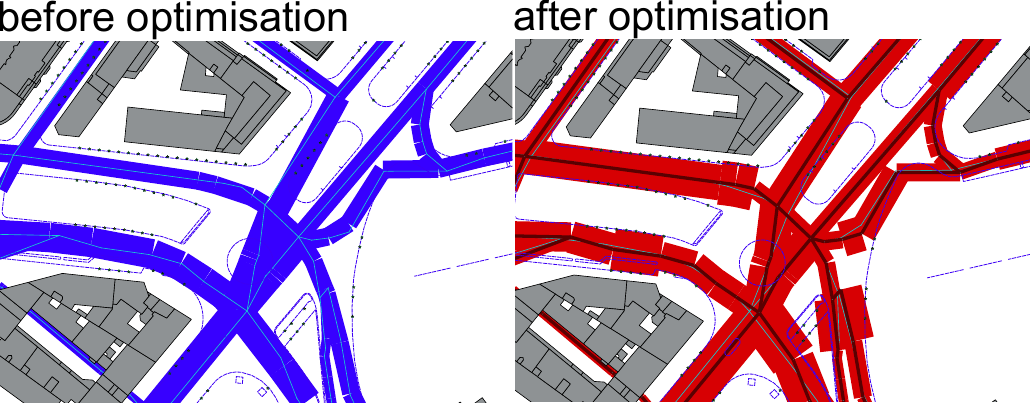}{Even in very complex situations, where observations are too sparse and the simple road model is not sufficient, results are stable (no constraints used).}{optim.fig.example_complex}{1} 
		
		We also looked at the impact of the regularisation constraints that limit changes (See Section \ref{optim.method.regularisation_force}).
		Clearly, constraints can degrade the result, leading to less correctly fitted roads (See Fig \ref{optim.fig.result_paris_constraints}).
		This phenomenon is particularly showing when looking at distance between ground-truth side-walk and road surface envelope (Table \ref{optim.tab.whole_paris}), especially looking at histograms (\ref{optim.fig.whole_area_histograms}).
		We stress that constraints can stille be introduced, but with an influence reduced at will using force weight. That way, the constraints weight can be adapted to the trust the user has in the observation and road model parameters initial values.
		
		\myimageHL{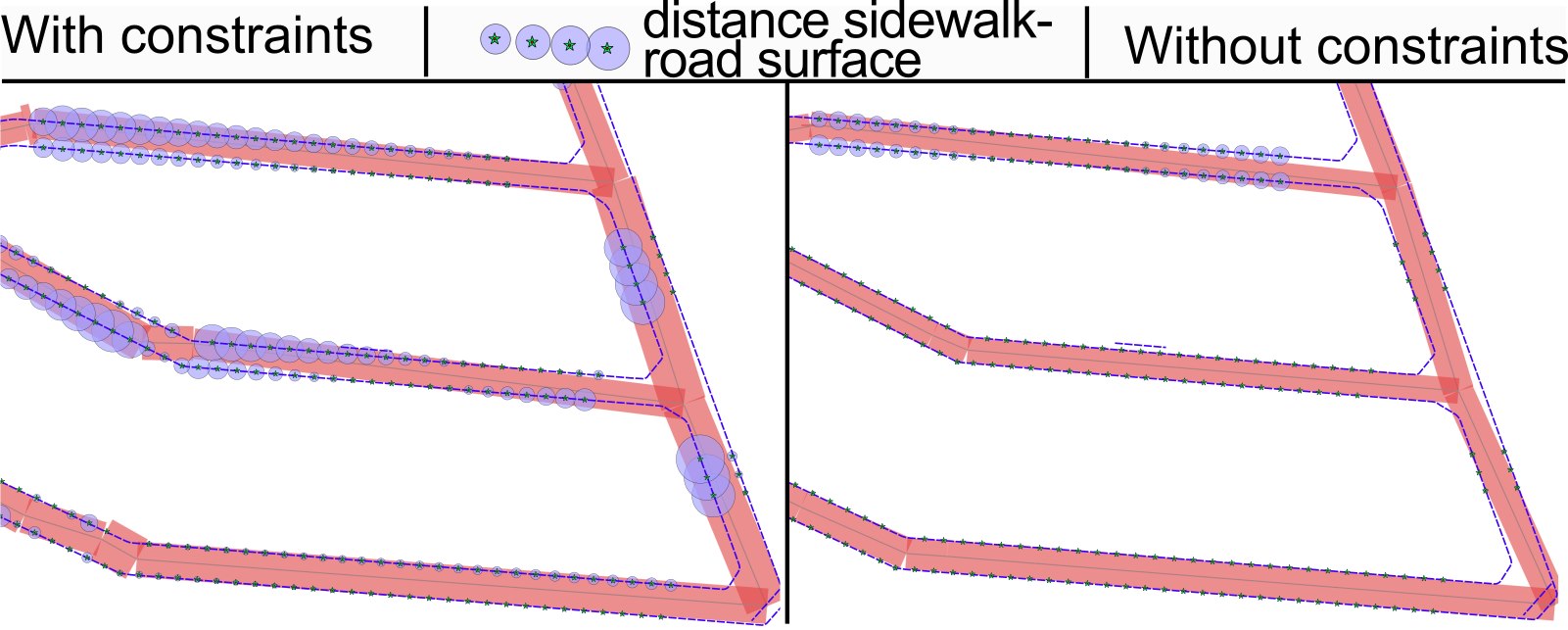}{Constraints are necessary to stay close to the initial solution. However, they can significantly degrade the results when there are enough quality observations. Circles are proportional to the distance between ground truth and result. Note that constraint influence can be modulated with weights.}{optim.fig.result_paris_constraints}{1}
		
		\begin{table}[h]
			\centering
			\caption{For whole Paris, distance to side-walk before and after optimisation using or not regularisation forces.}
			\label{optim.tab.whole_paris}
			\scriptsize 
			\begin{tabular}{cccc}
				Type & Mean (\metre) & Median (\metre)  &  Std dev (\metre) \\ 
				Initial (no optimisation)	& 	1.797 & 1.547 & 1.357 \\  
				Observation from ground truth (ODParis Sidewalk)	& 0.392 & 0.242 & 0.479 \\   
				
				Observation from ground truth, no regularisation	& 0.316 & 0.104 & 0.638 \\
			\end{tabular} 
		\end{table} 
		
		\myimageHL{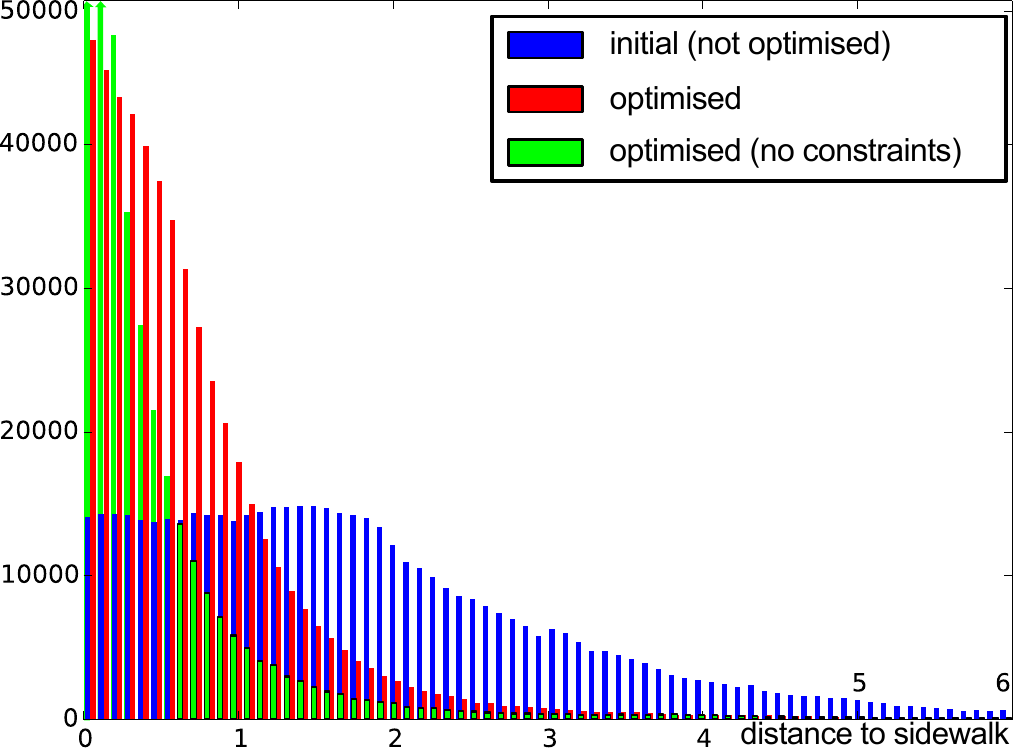}{Histogram of absolute distance to ground truth side-walk for "Paris" (optimisation using ground truth as observations).}{optim.fig.whole_area_histograms}{1} 
			
		\subsubsection{Results on "sensed" area.}
		The second area of experiment is the place where sensing data are available (See Sec. \vref{optim.result.experiment_areas}).
		
		Whatever the observation type used, the fitted model is always better than the initial road modelling,
		although using only object-observations improves only marginally the fitting.
		
		Results using kerb observations are subjectively good when enough observations are available.
		Quantitative results (Table \ref{optim.tab.sensing_area}) confirm this impression.

		However, there is a large difference between results obtained with observations from the ground truth (median = 0.12 \metre) and results obtained with kerb observations from sensing (median = 0.61 \metre). 
		The visualisation of observations and distance between ground truth side-walk and road model immediately informs about the problem (See Fig. \ref{optim.fig.result_whole_area}, bottom right illustration).
		Errors come mostly from sparsity of kerb observations (most often no observation, and in a few cases only one side of the road).
		Indeed, finding the correct road segment and road width is not possible if observations are only available on one side of the road, or not available!
		
		If the whole optimisation process were to not use any user inputs, using sensing data clearly improves fitting (median distance to ground truth is optimised from $ 1.51 \metre$ to $0.61 \metre$), but it still remains far from reachable results with more extensive observation coverage.

		Figure \ref{optim.fig.result_whole_area} gives an overview of results on whole sensed area.
		\myimageHL{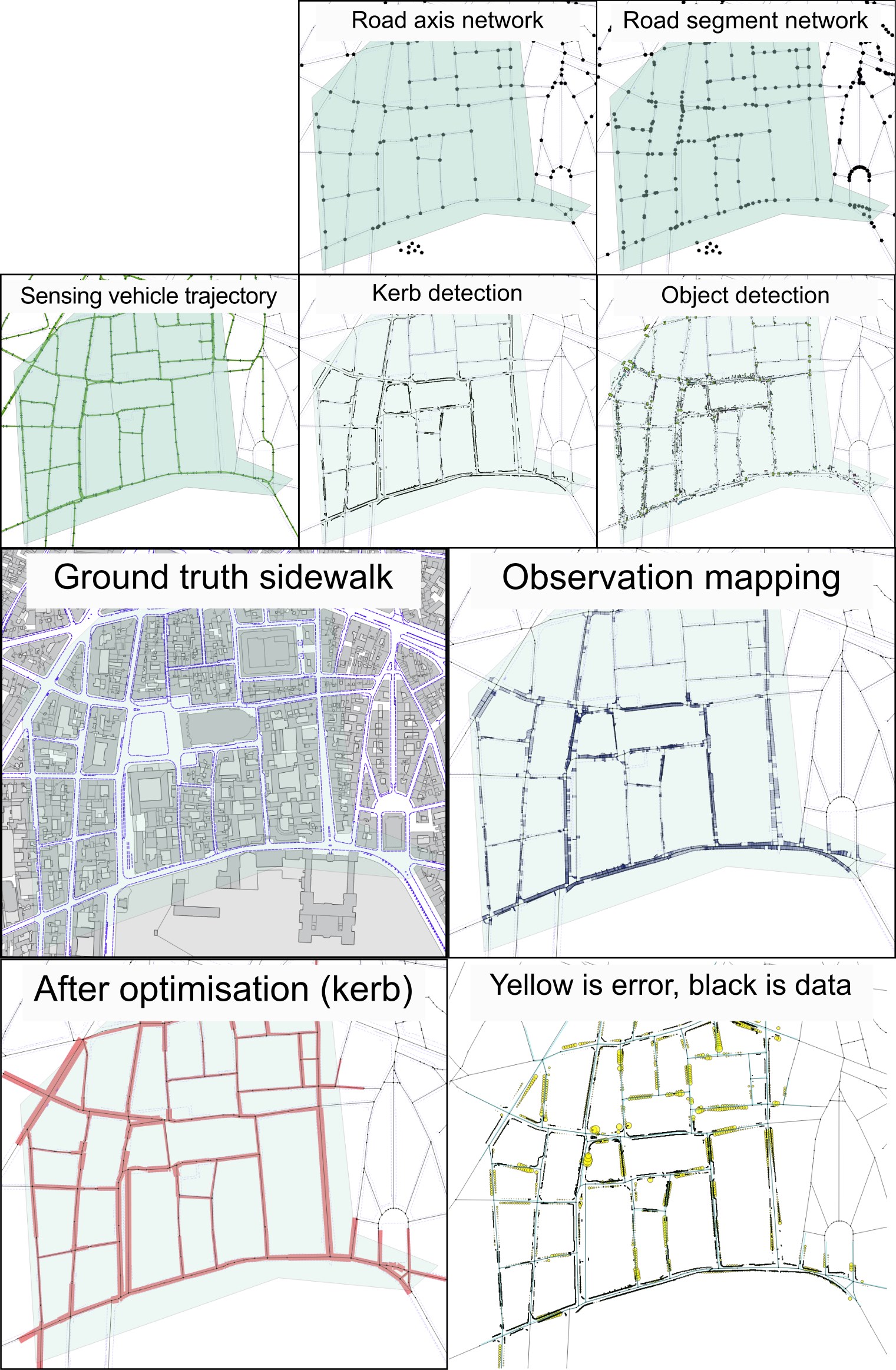}{Optimisation parameters are the street axis network and width attribute, which get split into a road segment network. A street sensing vehicle detects kerbs and objects, which are processed into observations, and matched to street segments. Non-linear least square optimisation fits the parameters to these observations.}{optim.fig.result_whole_area}{1}

	\begin{table*}[]
		\centering 
		\scriptsize 
		\begin{tabular}{cccc}
			Type & Mean (\metre) & Median (\metre)  &  Std dev (\metre) \\ 
			Initial (no optimisation)	& \textbf{1.856}  & \textbf{1.507}  & \textbf{1.534} \\ 
			Observations from ground truth (ODParis Sidewalk)	& 	0.481  &0.356  & 0.469 \\   
			Observations from ground truth, no regularisation	& 	0.281  & 0.118  & 0.442 \\ 
			Observations from ground truth, no regularisation, split road seg >10 \metre	& 	\textbf{0.117} & \textbf{0.014} & \textbf{0.306} \\ 
			Using Kerb observations	& 0.825  & 0.608  & 0.783 \\  
			Using Object observations	& 1.554 & 1.237 & 1.201  \\   
		\end{tabular} 
		\caption{For the entire sensing area, distance to side-walk before and after optimisation with diverse observations.}
		\label{optim.tab.sensing_area}
	\end{table*}

		\subsubsection{Results on "user" area}
	The last experiment area is a small part of the "sensing" area.
	We manually ensure that the road segment network is sufficiently divided.
	The initial road model (Fig. \ref{optim.fig.result_small_data}, top right, viewed with StreetGen) is quite far from the ground truth (top left), especially concerning road segments width.
	Observation matching (bottom left and right) shows the sparsity of observations (especially kerb) and noise (especially objects).
	
	\myimageHL{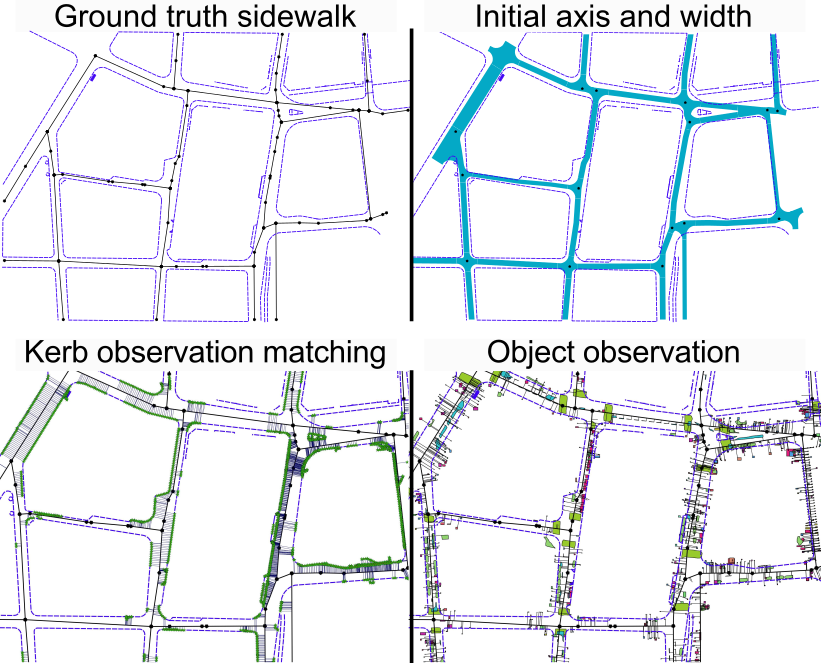}{Input data (road axis network + width with matched observations from sensing) for optimisation in the "user" experiment area. }{optim.fig.result_small_data}{1}
	
	Optimising using kerb observations produces a convincing fitting where observations are available (see Fig. \ref{optim.fig.result_small_kerb}, left, and Table \ref{optim.tab.small_sensing_area}).
	Because bad fitting happens when observations are missing, a user can input manually relatively few kerb points (middle).
	The resulting fitting after optimisation (Fig. \ref{optim.fig.result_small_kerb}, right), is extremely close to the ground truth.
	However, the distance to ground truth (Table \ref{optim.tab.sensing_area}, row 4) (median = 0.36 \metre) is still higher than the median distance reached for all of Paris with "perfect" observations.
	This resulting distance is then a sign of the limit of the expressiveness of the proposed road modelling (road segment + width).
	For instance, when the road width increases only on one side of the road.
	Indeed, the chosen "user" area contains many intersections, and problematic situations, like road with non constant width, and complex pedestrian crossings with pedestrian islands in the middle of the roadway.
	
	We further test the missing data hypothesis with an alternate kerb observation data set, with more coverage, but also much more noise.
	As expected, the results are better (Table\space\ref{optim.tab.small_sensing_area}), the additional noisiness is mostly dealt with by the loss function. 
	
	We also experiment using object observations (Fig. \ref{optim.fig.result_small_objects}). Object observations are less noisy than in other places,
	and so they can be used for optimisation on their own to obtain notably better fitting (Table \ref{optim.tab.small_sensing_area}), with median distance going from initial $1.8 \metre$ to $0.96 \metre$.
	However, results are still far from those obtained with kerb observations.
	
	To complete this, we use marking detection from \cite{Hervieu2015}, which are especially complete and with a limited noise (advanced markings).
	After fusion and optimisation, the result is significantly improved compared to other object observations (median distance of $0.78 \metre$ ).
	
	(See Discussion (Section \ref{optim.discussion.user_area}) for analyse of other results of table \ref{optim.tab.small_sensing_area}).
	
		\begin{table*}[h]
			\centering 
			\scriptsize 
			\begin{tabular}{cccc}
				Type & Mean (\metre) & Median (\metre)  &  Std dev (\metre) \\ 
				Initial (no optimisation)	& \textbf{2.044}  & \textbf{1.8}  & \textbf{0.937} \\  
				Ground truth (ODParis Sidewalk)	& 0.394  & 0.237  & 0.448 \\ 
				Ground truth (no constraint)	& \textbf{0.22}  & \textbf{0.084}  & \textbf{0.404} \\  
				Using Kerb observations	& 0.709 & 0.455 & 0.724 \\ 
				Using Kerb observations and User input	& 0.457 &0.34 & 0.413\\ 
				Using Kerb observation (alternate d.) & 0.496 &0.39 &0.413 \\
				Using Object observations	& 1.344 & 0.963 & 1.311  \\   
				Using Object observations (only markings)	& 1.53 & 1.169 & 1.272  \\   
				Using kerb and Object (cars,markings,signs) &	0.67 & 0.451 & 0.696 \\  
				Using advanced markings (only few streets) & 1.009&  0.782&  0.944 \\
			\end{tabular} 
			\caption{For a small part of sensing area, distance to side-walk before and after optimisation using diverse observations.}
			\label{optim.tab.small_sensing_area}
		\end{table*}

	\myimageHL{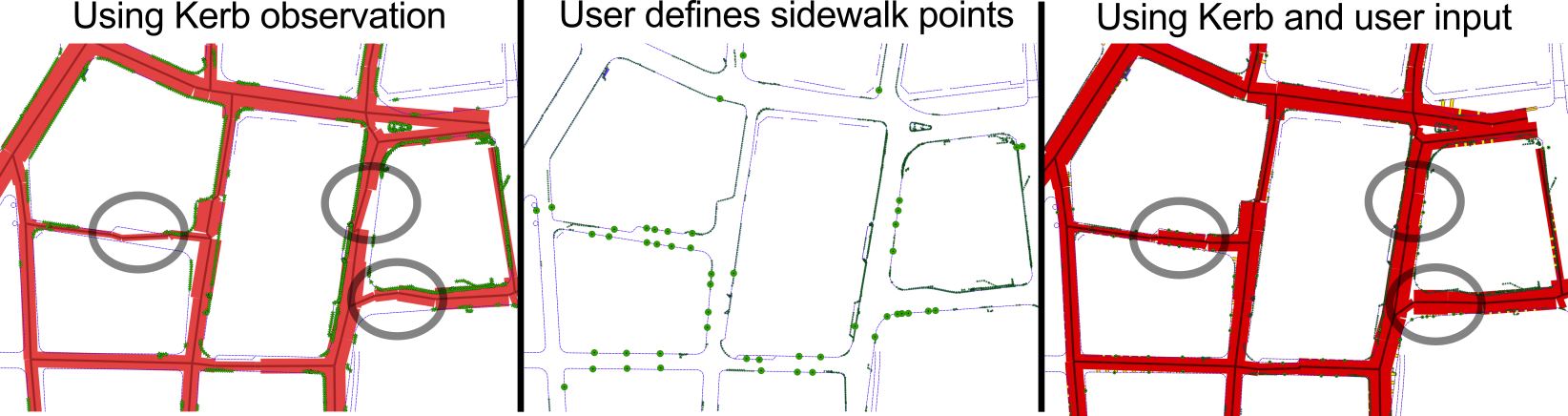}{Optimisation results based on kerb observations for "user" experiment area.}{optim.fig.result_small_kerb}{1}
	
	\myimageHL{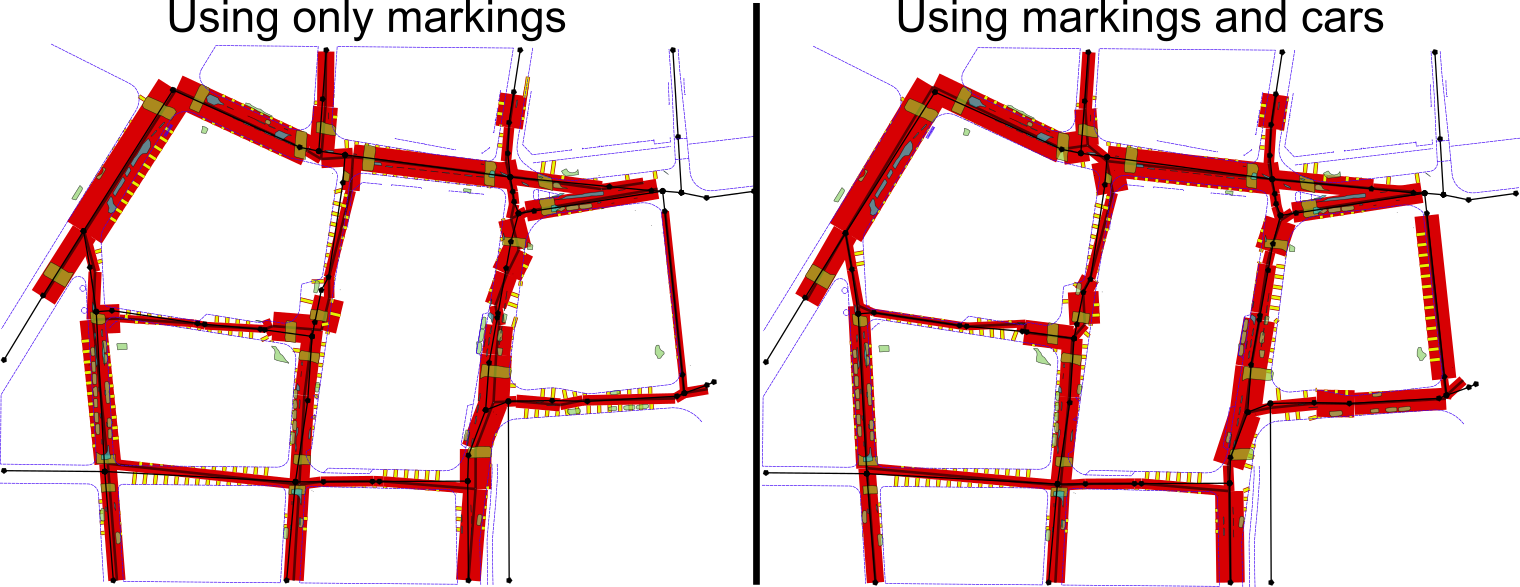}{Optimisation results based on object observation for the "user" experiment area.}{optim.fig.result_small_objects}{1}

	\subsection{Generating streets from optimised road model}
	In Section \vref{optim.method.optim_sg} we introduced how the optimised road network is used to generate a street network using StreetGen (See \cite{Cura2015a}).
	We test the method on two optimisation results: first the road modelling obtained on the "Sensing" experiment area using kerb observations, then the road modelling obtained on the "Paris" experiment area using the observations extracted from ground truth side-walk, with a road axis network split so no road segment is longer than 30 \metre.
	
	The process to regroup road segments is quite fast, with a few seconds for "Sensing" and a minute for "Paris". 
	See figure \vref{optim.fig.sg_optim} for illustration on result.
	Produced roads and intersection surfaces are available\footnote{\url{https://github.com/Remi-C/Network_snapping/tree/master/data/street_gen_after_optim}}.
	
	For "Paris" experiment area, the initial road axis network contains 19531 polylines, which are then split into 74325 road segments.
	After optimisation, those segments are regrouped into 37262 polylines.
	
	Optimisation does not guarantee to preserve topology (See Fig. \ref{optim.fig.error_topo}).
	We detect these errors (74) and display them, so an user can fix them.
	
	\myimageHL{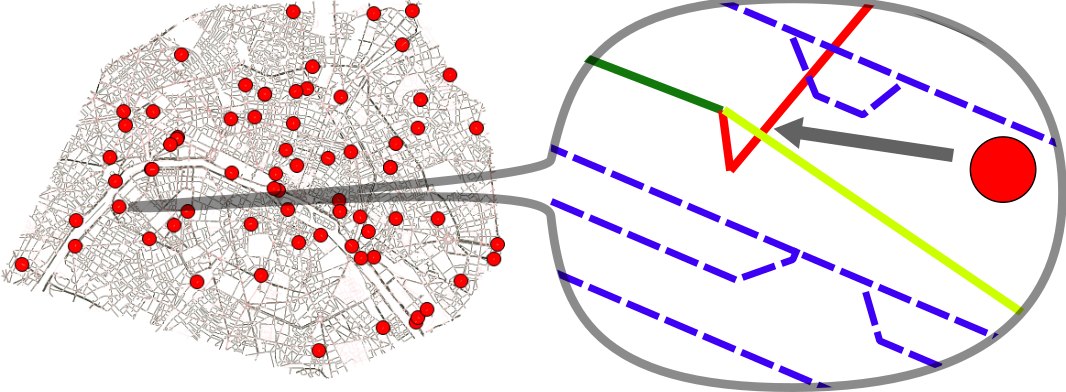}{After the optimisation process, the road axis network contains a few topological errors despite the use of constraint forces. These 74 errors for the whole "Paris" experiment area can be corrected by a user in about 10 minutes.}{optim.fig.error_topo}{1}



\section{Discussions }
	 \label{optim.discussion}  
	 This section discusses limitations and perspectives for key elements of our method and its results.
	 First, we discuss the choice of the road model, optimisation framework and forces choices.
	 Then we discuss how raw data are processed into observations, and how these observations are matched to road segments.
	 We then discuss the optimisation process and how to visualise its output.
	 Last, we discuss optimisation results.
	 
	\subsection{Modelling the problem}
		Our road modelling (road axis segment + width) suits well a city like Paris that has yet very complex roadway. 
		Our road model can not take into account road with non constant width, asymmetric road width change, pedestrian islands in the middle of a roadway, etc.\\
		In fact the current optimisation method could be extended in a minor way by introducing a road segment type with a linearly varying road width (possibly with variation different on each side).
		Such an extension would even be sufficient to model asymmetric road width change in most situations.
		This change would greatly enhance the capabilities of the road model.
		
		Another perspective would be to augment the road model with some "free-form" road sections that would be much less constrainted that regular road sections (for instance, any polygon). This could greatly improve the expressiveness of the road model, at the price of it being partially less abstract.
		
		The choice made toward robust non linear least square optimisation proves to be capable of fitting the chosen model to observations, and is reasonably robust to outliers in observations.
		Because it is fast, it is a good candidate to be integrated into a more powerful framework such as Reversible Jump Markov Chain Monte Carlo (RJMCMC, \cite{Green1995}) which can handle changes on numbers of parameters, extension of road model into different type of road segment, and possibly remove outliers.

	\subsection{Modelling observation effect as forces} 
		We consider the optimisation problem as a mechanical problem where observations produce forces on road segments and road width.
		
		\subsubsection{Observation forces}
		\paragraph{Kerb points}
			Forces induced by kerb points are extremely fast to compute thanks to Eigen\footnote{\url{http://eigen.tuxfamily.org}} (\cite{eigenweb}) library.
			These forces have a drawback: distance between observation and road segment is computed for an infinite road segment. It could be an improvement to actually check that the observation projection is inside the road segment, which would amount to disable some observations when road segment changes a lot.
		\paragraph{Street objects}
			Forces induced by road objects have a lot of room for improvement. 
			We use GEOS\footnote{\url{https://trac.osgeo.org/geos/}} to compute forces based on object observations, which greatly increases computing time (by a factor 10 to 100).
			We chose a sophisticated distance, based on the actual shared area.
			It involves computing distance between road segments and complex surfaces, and computing intersections between road segment surface and observation surfaces.
			The reason behind was to exploit exactly the data, without introducing approximation.
			However, there is no point to use such a precise measure when object detections and observations are so noisy! 
			Moreover, observations are systematically converted to surfaces.
			This is inefficient: an object observed as a point (a pole for instance) could use a point to segment distance.
			Considering the level of precision of the observations, GEOS could be abandoned to perform surface to segment distance and intersection area.
			Instead, all object observations could be modelised as points or rectangles, and distances and intersection could be directly computed with Eigen.
			
		\paragraph{Road segment azimuth from kerb points }
			Target road segment slope has \\
			proved to be successful at accelerating computing and obtaining better results.
			It may be due to the robust (although crude) way it is estimated (weighted median). 
			More advanced methods such as RANSAC could make it more robust. 
			Moreover, this force is sometimes meaningless for very short road segments that serve as transition in places where the road model is not sufficient compared to the road complexity.
			Such cases could be detected automatically, and slope force cancelled.
			User inputs are not included to compute this force, which could also be a significant improvement. 
			
		\subsubsection{Regularisation forces}
		All regularisation forces suffer from  a drawback: they are hard to set, because their value is a constant, whereas observation forces strength depends on the number of observation.
		Regularisation forces could be normalised regarding the number of observations, yet their current behaviour can be seen as desirable (a road with many observations will have low regularisation, and with few observation strong regularisation).
		
		\paragraph{New road width propagation force}
			We could improve results in case observations are missing by regularising the width given the neighbour segments width. 
			Let us take the example of a road axis divided in three road segments, with good observations on the first and last segments, but no observations for the middle segment. 
			First and last segments are moved and their widths changed so they fit the observations.
			Indirectly, the middle segment is also moved as its end nodes are shared with other segments.
			However, there is no such mechanism for width sharing.
			We could easily add a force that pushes toward having the same width as neighbour segments.
			This force would greatly increase results in place where the road axis network is over segmented.
			
			Similarly, we could add a force limitating the difference of width between two successive road segments.
			
		\paragraph{Distance to initial road segment node position} 
			
			We consider this force to be very important for scenarios where only a small part of the road network is to be modified.
			Indeed, it would be critical for user interaction that a change made by a user at a place (South of Paris for instance) does not produce changes far away from this place (North of Paris for instance).
			This force is a way to achieve that.
			Let us take the example where the user wants to edit one road axis.
			This force can be gradually stronger for edges as they are inside 1-neighbourhood, 2-neighbourhood, etc.
			
			In a similar scenario, this force is useful for pending edges that are directly connected to optimised edges but have no observations.
			
		\paragraph{Distance to initial road segment length}
			This force could also be modulated, as it is counter-productive for small segments that serve as transition whent he road model is not powerful enough.
		\paragraph{Distance to initial road segment width}
			This force works well to prevent isolated noisy observation to wrongly radically change road width.
			It proved to be particulary useful when dealing with noisy objects, as it prevents the width to increase out without sound reasons. 
		\paragraph{Distance to initial pair of road segment angle}
			We performed an approximation to compute this force due to lack of time.
			Yet this approximation may be ill suited when changes of angles are large, possibly leading this force to be inefficient or even counter productive.
			We possibly introduce an error in optimisation.
			The exact analytical change could be computed, which may lead to faster convergence, and maybe even  better results.
			
	\subsection{From raw data to observation} 
	 \paragraph{Street Lidar for kerb observations}
	 The main limitation of kerb observations is not the noise but the sparsity.
	 It has a much stronger effect on optimisation outcome than outliers.
	 Some streets have very few kerb detections, sometime on only one side of the street.
	 The essence of the problem is that side-walks are often masked by parked cars or street objects.
	 This is due to the used Lidar device (Riegl), which samples points in a plan orthogonal to the vehicle heading.
	 That way, a spot of side-walk may only be seen from one angle, which would then be easily occluded.
	 
	 We considered the interest of using another Velodyn Lidar as a complementary Lidar.
	 The Velodyn lidar is rotating, and thus can sense points from many different angles when the vehicle closes in and out (See Fig. \ref{optim.fig.velodyn}).
	 Figure \ref{optim.fig.velodyn} clearly shows that Velodyn greatly increases side-walk and kerb coverage. If the distance is sufficient, kerbs can even be sensed with ray passing under the cars.
	 
	 However, the Velodyn point cloud is also noisier and much larger. With increased data volumes and a need to fusion points, Velodyn raises practical challenges.
	 The main one is that as opposite to Riegl, point clouds can not be processed linearly, file by file. Indeed a portion of sidewalk may be seen by the Velodyn Lidar in many files, not necessary temporally close to each other. 
	 This problem is hard to solve with a file based solution, but would be quite easy to deal with the Point Cloud Server (See \cite{Cura2016a}).
	 
	 \myimageHL{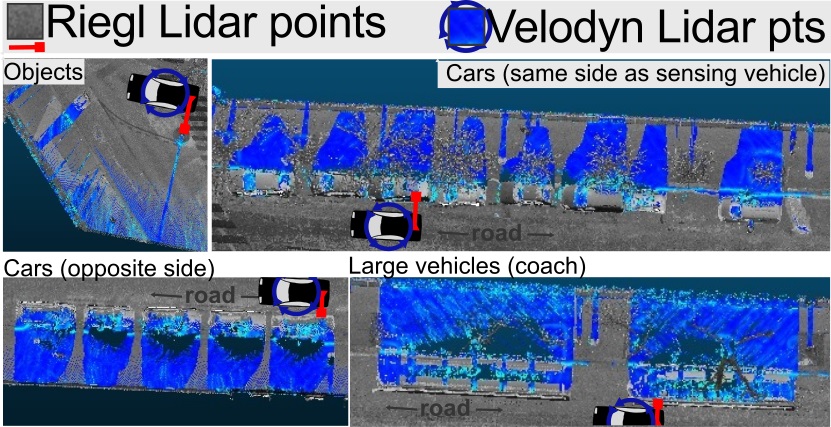}{Point cloud sensing with static Lidar (Riegl) is occluded by cars and street objects, which hides side-walks. Adding a rotating Lidar (Velodyn) greatly reduces occlusion because the same spot can be sensed from different angles.}{optim.fig.velodyn}{1}
	 \paragraph{Street Lidar for Object observations}
		 Objects produced by \cite{Serna2014} are most often too noisy to be used. 
		 In some cases, carefully filtered car observations can be used for rough road fitting (such as in "user" experiment area).
	 \paragraph{Markings}
	 We mainly use markings detected from stereo-vision (\cite{Soheilian2013}), with a low semantic content and only the main markings detected (pedestrian crossing and center lines). 
	 However, a more powerful new method (\cite{Hervieu2015}) has been shown to detect more diverse markings, with the potential to detect parking place markings. Those markings are important because they are close to the road border, and thus give a good estimate of road width.
	 Indeed, a central line marking is currently not very useful for the optimisation, as there are very few chances that this markings would be outside of the initial road surface.

	\subsection{Observation matching}
		\paragraph{Wrong matches}
		Intersections are ambiguous place for matching, partially because our model (road segment + width) is blatantly wrong in those places. 
		For kerb observations we simply exclude observation within intersection area.
		Yet we can not proceed similarly for object observation, as some objects are most frequently found in intersection area, such as pedestrian crossings and cars.
		
		The problem is that a large object may be at 0 distance of two road segment (i.e. intersection both).
		A a supplementary criteria to distance to implicit road surface could be added to disambiguate in this cases.
		The additional criteria could be a robust shared area measure, where the implicit road surface sharing the most area with the object to be mapped would be the one the object would be mapped with.
		Another solution would be to add object knowledge, such as orientation, and favor matching object with road having a similar orientation.
		Although it would be an easy change, we feel it potentially breaks the genericity of the input we use.
		
		\paragraph{Dynamic matching}
		The chosen optimisation solution is sensitive to outliers, be it real incorrect observations, or observations that are incorrectly matched.
		We stress that even if the matching is correct before optimisation, it may become wrong during optimisation when the edges are moved. 
		We detected several cases of such matching turning wrong.
		A solution could be to perform dynamic matching.
		The first way would be within the optimisation, for each iteration, observation generating large forces are re-mapped or deactivated.
		The second more discrete way would be to perform several successive optimisations, each time re-matching what looks like outliers, possibly removing them. 
		In this approach residual distribution of forces can be statistically analysed to find outliers,
		but it remains an adhoc process.
		
		\paragraph{Optimal matching}
		Another perspective would be to include the matching in a more powerful optimisation process.
		In this case, the system would optimize the parameter values, the number of parameters, as well as the matching between observation and parameters (including removal of outlying observations). This solution would be theoretically sound and could rely on good initialisation, but the design and more importantly balance between energy terms might be difficult, and the computing time potentially prohibitive.
	\subsection{Optimisation} 
		Several settings can be changed for optimising.
		Main settings are weight of each type of forces, and scale factor of loss function.
		In total, it amounts to less than $1O$ program-settings.
		We choose those settings with an empiric trial and error methodology, which is less than satisfactory.
		
		A ground truth road segment network and width could be manually entered for a small area, and then those program-settings could be found with a meta optimisation by measuring the distance to user-defined road model, either with stochastic method or with a similar non linear least square (although gradient would have to be obtained numerically and not analytically).
		
		We consider this to be essential to analyse the full potential of our optimisation method, as we could currently be using non optimal program-settings.
		
	%
		Ideally, we would not only find the best parameters for road axis segment position and width, but also be able to introduce new segments or delete unneeded ones.
		Similarly, parameters are changed to fit observations, yet some of this observations may be erroneous, and should be ignored.
		This would require an optimisation method that not only find the optimal values for parameters, but also find the optimal number of parameters, along with the optimal set of observations!
		Methods with this kind of capabilities are pretty limited to direct approach(brute force), more or less clever heuristics, and RJMCMC.
		In every case, we can see those kind of optimisation as meta optimisation, where one level find optimal parameters values for a given number of parameters, and a second level find the optimal number of parameters (the two levels are intertwined). 
		Therefore, without loss of generality, we focus on designing a method that find optimal parameters values first (first level).
		This method could be further used in more complex optimisation framework (second level).
	%
%
	\subsection{Results and Forces visualisation}
		Visualising results directly within a GIS (QGIS in our examples) is extremely handy, as results can be compared to aerial view and other reference vector layers.
		Moreover, the GIS software is necessary anyway for the user to input manual kerb point and/or correct automatic observations. 
		Currently the optimisation software is independent, but it could be integrated as a plug-in, or automatically triggered when user modifies the observations for instance.

	\subsection{Optimisation results}
	\subsubsection{Result evaluation}
		
		\paragraph{"Paris" experiment area}
		 The result (Table \ref{optim.tab.whole_paris}) at Paris scale ("Paris") with ground truth eliminate observation bias.
		 Without constraints, inputing ground-truth as observation, we get in the $0.1 \metre$ range from ground truth.
		 We explain this remaining distance by the limit of expressiveness of the road model (no asymetric width, no linear varying width), and more casually, by the ground truth own faults (large avenue ground truth contains side-walk but also other road features), and by overconstraints (road axis are not split enough).
		 Figure \ref{optim.fig.result_limitation} illustrates those common limitations.

		\myimageFullPageWidth{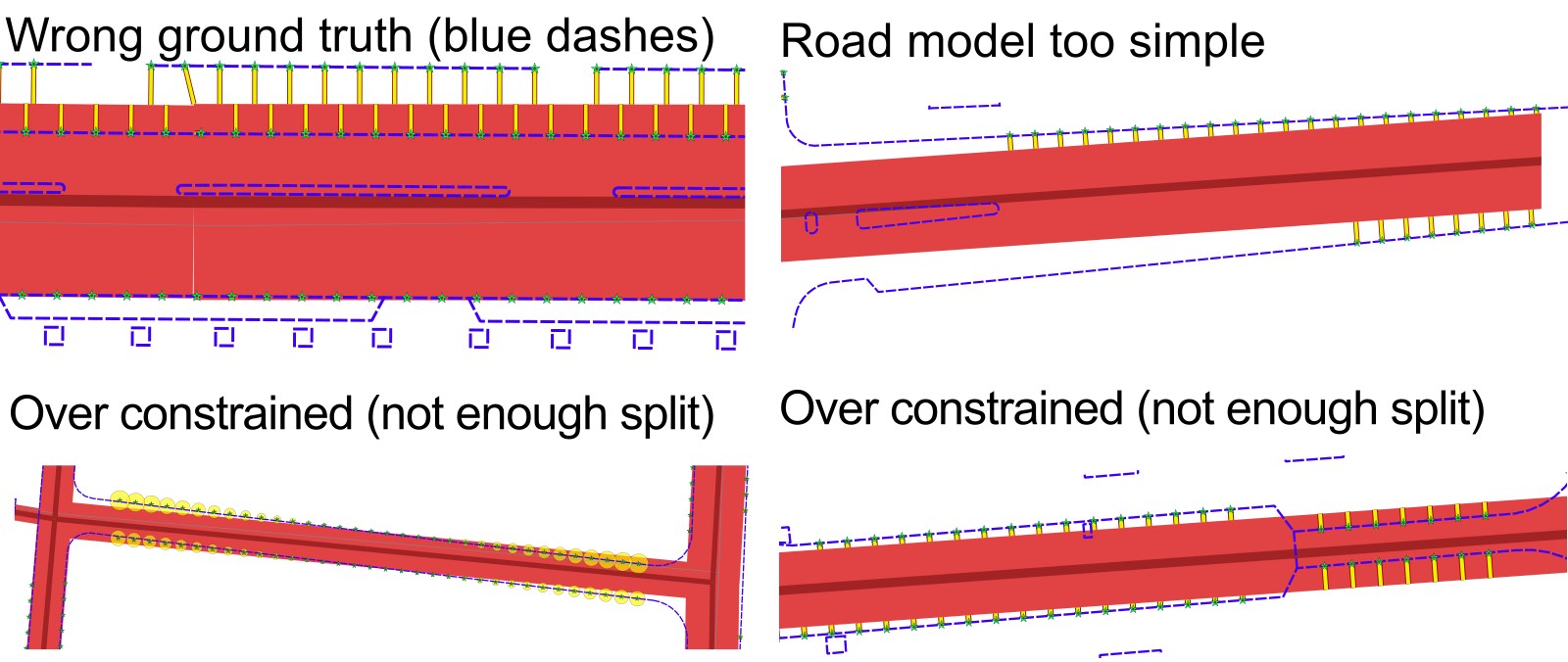}{Various illustrations of optimisation process limitation. The data used as ground truth contains errors. Then road axis main not split enough. Last, the simple road model used can not deal with some features of Paris roadway.}{optim.fig.result_limitation}
		 
		 All factors included, $0.1 \metre$ is a very satisfactory result, as it would be about the same size as a precise aerial image pixel, or about Open Data Paris official precision.
		 
		 Using forces to resist change and stay close to the initial road model is unnecessary when using reliable and dense observations.
		 We could not find a place where regularisation forces were really useful.
		 In a similar topic, optimisation appeared to be robust, even when the road model was obviously not suited to model the complex roads.
		 
		 Of course we cannot expect to have access to observations with such a high quality and coverage for a real life use case.
		 
		\paragraph{"Sensing" experiment area}
		Results (Table \ref{optim.tab.sensing_area}) on both regular and alternate dataset confirm that a fully automatic process with kerb observation can reach $0.6 \metre$ to ground truth distance, thus significantly improving the initial median distance ($1.5 \metre$), even using quite noisy observations.
		The main reasons explaining the difference between this result and the result using observation from ground-truth is visually immediate: observation sparsity.
		
		Given the object detection we dispose of, they are either too sparse (road signs, road markings) or too noisy (cars, other objects) to be really useful in a full automated process. 
		They only marginally improve model fitting, at a high computational cost.
		However such object observations have a real potential, as proved by a simple experiment we performed.
		We manually created a complete marking dataset for a road. After optimisation, the model was a really good fit (parking place markings are especially useful, as well as correct pedestrian crossings).
		Similarly, markings from \cite{Hervieu2015} which are more complete delivered promising results.
		
		"Sensing" area is fairly sized, making it representative. 
		For the goal of evaluating the impact of road axis being over-constrained (not enough split),
		we automatically split the road axis so that no road segment is more than $10 \metre$ long.
		This gives a fairly good certitude that road axis are no more over-constrained.
		The average distance to ground truth after optimising using ground truth observations is about $0.1 \metre$ on \emph{average}, with a median distance of $0.015 \metre$, without regularisation forces and excluding observation in intersection area (\metre) from our measures.
		Therefore, the only remaining factor explaining this residual is the fact that our road model is too simple to explain complex Paris roads.
		All considering, this proves that even the very simple road model is sufficient to come very close to complex road surface of Paris.
		
		\paragraph{"User" experiment area}
		\label{optim.discussion.user_area}
		The reduced area is especially telling regarding road model limitation (Table \ref{optim.tab.small_sensing_area}), although this area is smaller so less representative.
		In this area, we can  manually control that road axis are split into the correct number of segments, and that sidewalk ground truth only contains sidewalk.
		When using observation from ground truth, the only error factor remaining is then the road model limitations (See Fig. \ref{optim.fig.errors_sources}).
		Despite some very complex road features (asymmetric width, pedestrian island, non constant road),
		the model is very close to ground truth ($0.1 \metre$).
		
		Using sensing data on the area yields an interesting result, as it gives a realistic example of automatic fitting for a well split road axis network.
		Optimisation results are about $0.45 \metre$ from ground truth side-walk.
		For this area we also tested a user-assisted approach, where user complement sensed kerb observation with manually added kerb points. 
		The result is visually more acceptable, but quantitatively not very different.
		Indeed, results in this range hit the sensing vehicle geospatial precision, which is about $0.4 \metre$ in this area where streets are narrow and buildings are high (masking GPS).
		
		Experiments using object observations confirm that it would be hard to produce quality results using them.
		However kerb observations can be marginally complemented by object observations for a small diminution of average distance to ground truth.
	
\subsection{Generating streets from optimised road model}
		
		Before generating streets with StreetGen, we regroup road segments with similar width and assign probable width to road segments that have no observations.
		In a sense this amount to a regularisation of road widths, and it should be performed directly during the optimisation process, using for instance the proposed width regularisation force.
		
		StreetGen road model uses turning radius in intersection. These radius could also be determined by the optimisation process.
		
		Concerning the topology errors, forces could be introduced within the optimisation process so a road segment can not intersect another one (except 1-neighbours). However such forces would have to be limited to N-neighbours, with a small N (for instance 3), so the number of forces is not too large.


\section{Conclusion} 
	
	Our goal was to find a way to model city roads so it better fits various observations (kerb, street objects, etc.) in a city surrounding, at various scale (whole city (e.g. Paris), few hundreds axis, few blocks), starting from the approximate road axis network with coarse road width.
	 The road is modelled as a network of road segments with their own width.
	 We formulate the problem as an optimisation problem aiming at finding the best road parameters (position of segment node, width values) that fits the observations.

	Observations can be obtained automatically through sensing (street Lidar and images, aerial images), manually (user entered) and/or from legacy data (GIS vector layer describing side-walk).
	Observations are either consolidated kerb points detections, or street object detections, where each object type expected behaviour regarding the road is a setting (for instance pedestrian crossing markings are expected to be inside the roadway surface).
	We consider the optimisation problem similarly to a mechanic optimisation problem, with observation exerting forces on road segments and width, and other opposing forces trying to keep close to initialisation values.
	
	Finding the global optimal model to fit the observations is a complex task, as not only the position of road segments and their width have to be optimised, but also which observation affects what segment, and even the number of segments (creating/deleting/splitting/merging segments).
	In this work, we solve only a part of the optimisation problem within a non-linear least square framework, to find the road edge position and width optimal values. 
	Initial model fitting is not very good, although reasonable (median distance between model and (not perfect) ground-truth side-walk is $1.8 \metre$).

	The results with synthetic data (observation derived from ground truth side-walk) suggest that the optimisation process can produce a solution closely fitting the actual roads (median distance of $0.1 \metre$) at large scale quickly (e.g. whole Paris city in few minutes).
	This proves that even a simple road model is sufficient to model the vast majority of the complex Paris roadway.
	
	Fully automatic results with real sensing data show that kerb detection can be used effectively to fit the model (median distance 0.6 \metre), especially when the road network segments are adequately split (median distance 0.45 \metre).
	The main factor explaining the remaining distance seems to be the lack of data, which can be complemented by user input (median distance of 0.35 \metre), and the inherent sensing data precision (about 0.3 \metre).
	Object observations have to be reasonably noisy or they are not as useful as kerb observation (median distance of 1 \metre), even if they can be used to slightly complement kerb observations.
	
	Furthermore, our optimisation framework is fast enough (< 1 \second)to be used in an interactive guided scenario when focusing on few roads (for editing or corrections for instance).
	
	Promising and fast results make the proposed method suited to be integrated as a step into a more powerful optimisation framework, such as a Reversible Jump Markov Chain Monte Carlo method (RJMCMC, \cite{Green1995}), which would be able to fully deal with observation outliers removal and adequate introduction/removal of new road segments, as well as different type of road segment (including one with linearly varying width).

	\section{Acknowledgment}
	This article is an extract of \cite{Cura2016thesis} (chap. 5). We thank Prof.Peter Van Oosterom and Prof.Christian Heipke for their extensive review.


	\section{Bibliography}  
	\bibliography{./all_bibli} 

\begin{thebibliography}{xx}

\bibitem[Agarwal et al., 2016]{Agarwal2016}
Agarwal, S., Mierle, K. and {Others}, 2016.
\newblock Ceres {{Solver}}.

\bibitem[Ahmed et al., 2014]{Ahmed2014}
Ahmed, M., Karagiorgou, S., Pfoser, D. and Wenk, C., 2014.
\newblock A comparison and evaluation of map construction algorithms using
  vehicle tracking data.
\newblock Geoinformatica 19(3), pp.~601--632.

\bibitem[Aichholzer et al., 1996]{Aichholzer1996}
Aichholzer, O., Aurenhammer, F., Alberts, D. and G{\"a}rtner, B., 1996.
\newblock A {{Novel Type}} of {{Skeleton}} for {{Polygons}}.
\newblock In: H.~Maurer, C.~Calude and A.~Salomaa (eds), J.{{UCS The Journal}}
  of {{Universal Computer Science}}, {Springer Berlin Heidelberg}, Berlin,
  Heidelberg, pp.~752--761.

\bibitem[Airault et al., 1994]{Airault1994}
Airault, S., Ruskone, R. and Jamet, O., 1994.
\newblock Road detection from aerial images: A cooperation between local and
  global methods.
\newblock In: Satellite {{Remote Sensing}}, {International Society for Optics
  and Photonics}, pp.~508--518.

\bibitem[Bar~Hillel et al., 2012]{BarHillel2012}
Bar~Hillel, A., Lerner, R., Levi, D. and Raz, G., 2012.
\newblock Recent progress in road and lane detection: A survey.
\newblock Machine Vision and Applications.

\bibitem[Baumgartner et al., 1999]{Baumgartner1999}
Baumgartner, A., Steger, C., Mayer, H., Eckstein, W. and Ebner, H., 1999.
\newblock Automatic road extraction based on multi-scale, grouping, and
  context.
\newblock Photogrammetric Engineering and Remote Sensing 65, pp.~777--786.

\bibitem[Boyko and Funkhouser, 2011]{Boyko2011}
Boyko, A. and Funkhouser, T., 2011.
\newblock Extracting roads from dense point clouds in large scale urban
  environment.
\newblock ISPRS Journal of Photogrammetry and Remote Sensing 66(6),
  pp.~S2--S12.

\bibitem[Clode et al., 2007]{Clode2007}
Clode, S., Rottensteiner, F., Kootsookos, P. and Zelniker, E., 2007.
\newblock Detection and vectorization of roads from lidar data.
\newblock Photogrammetric Engineering \& Remote Sensing 73(5), pp.~517--535.

\bibitem[Cura, 2016a]{Cura2016thesis}
Cura, R., 2016a.
\newblock Inverse procedural Street Modelling : from interactive to automatic
  reconstruction.
\newblock PhD thesis, Universite Paris Est.

\bibitem[Cura, 2016b]{Cura2016a}
Cura, R., 2016b.
\newblock A scalable and multi-purpose {{Point Cloud Server}} ({{PCS}}) for
  easier and faster point management and processing.
\newblock ISPRS Journal of Photogrammetry and Remote Sensing.

\bibitem[Cura et al., 2015a]{Cura2015}
Cura, R., Perret, J. and Paparoditis, N., 2015a.
\newblock Point {{Cloud Server}} (pcs): {{Point Clouds In}}-{{Base Management}}
  and {{Processing}}.
\newblock ISPRS Annals of Photogrammetry, Remote Sensing and Spatial
  Information Sciences 1, pp.~531--539.

\bibitem[Cura et al., 2015b]{Cura2015a}
Cura, R., Perret, J. and Paparoditis, N., 2015b.
\newblock {{STREETGEN}}: {{IN}}-{{BASE PROCEDURAL}}-{{BASED ROAD GENERATION}}.
\newblock ISPRS Annals of Photogrammetry, Remote Sensing and Spatial
  Information Sciences II-3/W5, pp.~409--416.

\bibitem[Ester et al., 1996]{Ester1996}
Ester, M., Kriegel, H.-p., S, J. and Xu, X., 1996.
\newblock A density-based algorithm for discovering clusters in large spatial
  databases with noise.
\newblock proceedings of 2nd International Conference on Knowledge Discovery
  and Data Mining, {AAAI Press}, pp.~226--231.

\bibitem[Fischler et al., 1981]{Fischler1981}
Fischler, M.~A., Tenenbaum, J.~M. and Wolf, H.~C., 1981.
\newblock Detection of roads and linear structures in low-resolution aerial
  imagery using a multisource knowledge integration technique.
\newblock Computer graphics and image processing 15(3), pp.~201--223.

\bibitem[Green, 1995]{Green1995}
Green, P.~J., 1995.
\newblock Reversible jump {{Markov}} chain {{Monte Carlo}} computation and
  {{Bayesian}} model determination.
\newblock Biometrika 82(4), pp.~711--732.

\bibitem[Guennebaud et al., 2010]{eigenweb}
Guennebaud, G., Jacob, B. and {others}, 2010.
\newblock Eigen v3 software.

\bibitem[Hatger and Brenner, 2003]{Hatger2003}
Hatger, C. and Brenner, C., 2003.
\newblock Extraction of road geometry parameters from laser scanning and
  existing databases.
\newblock International Archives of Photogrammetry, Remote Sensing and Spatial
  Information Sciences 34(3/W13), pp.~225--230.

\bibitem[Hervieu et al., 2015]{Hervieu2015}
Hervieu, A., Soheilian, B. and Br{\'e}dif, M., 2015.
\newblock Road {{Marking Extraction Using}} a model\&data driven
  {{Rj}}-{{Mcmc}}.
\newblock ISPRS Annals of the Photogrammetry, Remote Sensing and Spatial
  Information Sciences 2(3), pp.~47.

\bibitem[McKeown and Denlinger, 1988]{McKeown1988}
McKeown, D.~M. and Denlinger, J.~L., 1988.
\newblock Cooperative methods for road tracking in aerial imagery.
\newblock In: Computer {{Vision}} and {{Pattern Recognition}}, 1988.
  {{Proceedings CVPR}}'88., {{Computer Society Conference}} on, {IEEE},
  pp.~662--672.

\bibitem[Montoya-Zegarra et al., 2014]{Montoya-Zegarra2014}
Montoya-Zegarra, J.~A., Wegner, J.~D., 'ubor Ladicky, L. and Schindler, K.,
  2014.
\newblock Mind the gap: Modeling local and global context in (road) networks.
\newblock In: German {{Conference}} on {{Pattern Recognition}} ({{GCPR}}),
  {Springer}, pp.~212--223.

\bibitem[Paparoditis et al., 2012]{Paparoditis2012}
Paparoditis, N., Papelard, J.-P., Cannelle, B., Devaux, A., Soheilian, B.,
  David, N. and Houzay, E., 2012.
\newblock Stereopolis {{II}}: {{A}} multi-purpose and multi-sensor {{3D}}
  mobile mapping system for street visualisation and {{3D}} metrology.
\newblock Revue fran{\c c}aise de photogramm{\'e}trie et de
  t{\'e}l{\'e}d{\'e}tection 200(1), pp.~69--79.

\bibitem[Quackenbush et al., 2013]{Quackenbush2013}
Quackenbush, L.~J., Im, I. and Zuo, Y., 2013.
\newblock Road extraction: A review of {{LiDAR}}-focused studies.
\newblock Remote Sensing of Natural Resources pp.~155--169.

\bibitem[Ravanbakhsh et al., 2008]{Ravanbakhsh2008}
Ravanbakhsh, M., Heipke, C. and Pakzad, K., 2008.
\newblock Road junction extraction from high-resolution aerial imagery.
\newblock The Photogrammetric Record 23(124), pp.~405--423.

\bibitem[Roeth et al., 2016]{Roeth2016}
Roeth, O., Zaum, D. and Brenner, C., 2016.
\newblock Road network reconstruction using reversible jump {{MCMC}} simulated
  annealing based on vehicle trajectories from fleet measurements.
\newblock {IEEE}, pp.~194--201.

\bibitem[Serna and Marcotegui, 2014]{Serna2014}
Serna, A. and Marcotegui, B., 2014.
\newblock Detection, segmentation and classification of {{3D}} urban objects
  using mathematical morphology and supervised learning.
\newblock ISPRS Journal of Photogrammetry and Remote Sensing p.~34.

\bibitem[Soheilian et al., 2010]{Soheilian2010}
Soheilian, B., Paparoditis, N. and Boldo, D., 2010.
\newblock {{3D}} road marking reconstruction from street-level calibrated
  stereo pairs.
\newblock ISPRS Journal of Photogrammetry and Remote Sensing 65(4),
  pp.~347--359.

\bibitem[Soheilian et al., 2013a]{Soheilian2013a}
Soheilian, B., Paparoditis, N. and Vallet, B., 2013a.
\newblock Detection and {{3D}} reconstruction of traffic signs from multiple
  view color images.
\newblock ISPRS Journal of Photogrammetry and Remote Sensing 77, pp.~1--20.

\bibitem[Soheilian et al., 2013b]{Soheilian2013}
Soheilian, B., Tournaire, O., Paparoditis, N., Vallet, B. and Papelard, J.-P.,
  2013b.
\newblock Generation of an integrated {{3D}} city model with visual landmarks
  for autonomous navigation in dense urban areas.
\newblock In: 2013 {{IEEE Intelligent Vehicles Symposium}} ({{IV}}),
  pp.~304--309.

\bibitem[Tupin et al., 1998]{Tupin1998}
Tupin, F., Maitre, H., Mangin, J.-F., Nicolas, J.-M. and Pechersky, E., 1998.
\newblock Detection of linear features in {{SAR}} images: Application to road
  network extraction.
\newblock IEEE transactions on geoscience and remote sensing 36(2),
  pp.~434--453.

\bibitem[Yang et al., 2013]{Yang2013a}
Yang, B., Fang, L. and Li, J., 2013.
\newblock Semi-automated extraction and delineation of {{3D}} roads of street
  scene from mobile laser scanning point clouds.
\newblock ISPRS Journal of Photogrammetry and Remote Sensing 79, pp.~80--93.

\bibitem[Zhang et al., 2009]{Zhang2009a}
Zhang, G., Zheng, N., Cui, C., Yan, Y. and Yuan, Z., 2009.
\newblock An efficient road detection method in noisy urban environment.
\newblock In: Intelligent {{Vehicles Symposium}}, 2009 {{IEEE}}, {IEEE},
  pp.~556--561.

\bibitem[Zhang et al., 2010]{Zhang2010b}
Zhang, L., Thiemann, F. and Sester, M., 2010.
\newblock Integration of {{GPS}} traces with road map.
\newblock In: Proceedings of the Second International Workshop on Computational
  Transportation Science, {ACM}, pp.~17--22.

\bibitem[Zhang, 2010]{Zhang2010a}
Zhang, W., 2010.
\newblock {{LIDAR}}-based road and road-edge detection.
\newblock In: Intelligent {{Vehicles Symposium}} ({{IV}}), 2010 {{IEEE}},
  {IEEE}, pp.~845--848.

\bibitem[Ziems et al., 2007]{Ziems2007}
Ziems, M., Gerke, M. and Heipke, C., 2007.
\newblock Automatic road extraction from remote sensing imagery incorporating
  prior information and colour segmentation.
\newblock IntArchPhRS (36), PIA 7, pp.~141--147.

\end{thebibliography}

%
%
%
%

\end{document}